\documentclass{elsarticle}  

\usepackage{lineno}
\usepackage{hyperref}

\usepackage[utf8]{inputenc} 
\usepackage[T1]{fontenc}    
\usepackage{url}            
\usepackage{booktabs}       
\usepackage{amsfonts}       
\usepackage{nicefrac}       
\usepackage{microtype}      
\usepackage{caption}        
\usepackage{subcaption}     
\usepackage{amsmath}        
\usepackage[capitalise]{cleveref}       
\usepackage{graphicx}       
\usepackage[toc, acronym, automake=false, nopostdot]{glossaries}
\usepackage{bm}             
\usepackage{tikz}           
\usepackage{multirow}       
\usepackage[frozencache,cachedir=.]{minted}         
\usepackage{xspace}         
\usepackage[binary-units=true]{siunitx}        
\usepackage{soul}           

\glsdisablehyper

\hypersetup{pdfauthor=author}

\newcommand{\torchioversion}{\texttt{v0.18.0}\xspace}
\newcommand{\myhref}[1]{\href{#1}{\texttt{#1}}}
\newcommand{\fnurl}[1]{\footnote{\myhref{#1}}}

\newcommand{\trsfl}[2]{\texttt{#1} \includegraphics[height=1.5\fontcharht\font`\B]{#2}}
\newcommand{\logo}[1]{\includegraphics[height=8pt]{#1}}
\newcommand{\comment}[1]{}

\definecolor{pytorch_orange}{HTML}{F84C36}
\definecolor{rev1}{HTML}{000000}  
\definecolor{rev2}{HTML}{000000}  

\modulolinenumbers[5]

\journal{Computer Methods and Programs in Biomedicine}

\graphicspath{{figures/}}

\begin{document}
    \begin{frontmatter}

    \title{%
        TorchIO: A Python library for efficient loading, preprocessing,
        augmentation and patch-based sampling of medical images in deep learning
    }

    \author[ucl,weiss,kcl]{Fernando Pérez-García\corref{correspondingauthor}}
\cortext[correspondingauthor]{Corresponding author}
\ead{fernando.perezgarcia.17@ucl.ac.uk}

\author[kcl]{Rachel Sparks}

\author[kcl]{Sébastien Ourselin}

\address[ucl]{Department of Medical Physics and Biomedical Engineering, University College London, UK}
\address[weiss]{Wellcome / EPSRC Centre for Interventional and Surgical Sciences (WEISS), University College London, UK}
\address[kcl]{School of Biomedical Engineering \& Imaging Sciences (BMEIS), King's College London, UK}

    \newacronym{api}{API}{application programming interface}
\newacronym{cli}{CLI}{command-line interface}
\newacronym{cnn}{CNN}{convolutional neural network}
\newacronym{copd}{COPD}{chronic obstructive pulmonary disease}
\newacronym{cpu}{CPU}{central processing unit}
\newacronym{ct}{CT}{computerized tomography}
\newacronym{cvpr}{CVPR}{Computer Vision and Pattern Recognition}
\newacronym{dag}{DAG}{directed acyclic graph}
\newacronym{dicom}{DICOM}{Data Imaging and Communications in Medicine}
\newacronym{dltk}{DLTK}{Deep Learning Toolkit}
\newacronym{dmri}{dMRI}{diffusion MRI}
\newacronym{fft}{FFT}{Fast Fourier Transform}
\newacronym{fmri}{fMRI}{functional MRI}
\newacronym{fov}{FOV}{field of view}
\newacronym{gif}{GIF}{geodesical information flows}
\newacronym{gmm}{GMM}{Gaussian mixture model}
\newacronym{gpu}{GPU}{graphics processing unit}
\newacronym{gui}{GUI}{graphical user interface}
\newacronym{ixi}{IXI}{Information eXtraction from Images}
\newacronym{minc}{MINC}{Medical Imaging NetCDF}
\newacronym{mni}{MNI}{Montreal Neurological Institute}
\newacronym{monai}{MONAI}{the Medical Open Network for AI}
\newacronym{mri}{MRI}{magnetic resonance imaging}
\newacronym{nifti}{NIfTI}{Neuroimaging Informatics Technology Initiative}
\newacronym{nips}{NeurIPS}{Neural Information Processing Systems}
\newacronym{nrrd}{NRRD}{Nearly Raw Raster Data}
\newacronym{pip}{PIP}{Pip Installs Packages}
\newacronym{png}{PNG}{Portable Network Graphics}
\newacronym{pypi}{PyPI}{Python Package Index}
\newacronym{rgb}{RGB}{red-green-blue}
\newacronym{rng}{RNG}{random number generator}
\newacronym{us}{US}{ultrasound}

    \begin{abstract}
        \paragraph{Background and Objective}

Processing of medical images such as \acrshort{mri} or \acrshort{ct} presents
different challenges compared to \acrshort{rgb} images typically used in computer
vision.
These include
a lack of labels for large datasets,
high computational costs,
and the need of metadata to describe the physical properties of voxels.
Data augmentation is used to artificially increase the size of the training
datasets.
Training with image subvolumes or patches decreases the need for computational
power.
Spatial metadata needs to be carefully taken into account in order to ensure a
correct alignment and orientation of volumes.

\paragraph{Methods}

We present TorchIO, an open-source Python library to enable efficient loading,
preprocessing, augmentation and patch-based sampling of medical images
for deep learning.
TorchIO follows the style of PyTorch and integrates standard medical image
processing libraries to efficiently process
images during training of neural networks.
TorchIO transforms can be easily composed, reproduced, traced and
extended.
Most transforms can be inverted, making the library suitable for
test-time augmentation and estimation of aleatoric uncertainty in the
context of segmentation.
We provide multiple generic preprocessing and augmentation operations as well as
simulation of \acrshort{mri}-specific artifacts.

\paragraph{Results}

Source code, comprehensive tutorials and extensive documentation for TorchIO can
be found at \myhref{https://torchio.rtfd.io/}.
The package can be installed from the Python Package Index (PyPI) running
\texttt{pip install torchio}.
It includes a command-line interface which allows users to apply
transforms to image files without using Python.
Additionally, we provide a graphical user interface within a TorchIO extension in 3D Slicer to visualize the effects of transforms.

\paragraph{Conclusions}

TorchIO was developed to help researchers standardize
medical image processing pipelines and allow them to focus on the deep learning
experiments.
It encourages good open-science practices, as it supports experiment
reproducibility and is version-controlled so that the software can be cited
precisely.
Due to its modularity, the library is compatible with other frameworks for
deep learning with medical images.

    \end{abstract}

    \glsresetall  


\end{frontmatter}

    \section{Introduction}

Recently, deep learning has become a ubiquitous research approach for solving
image understanding and analysis problems.
\comment{
This is in part due to increases in
computational power of \glspl{cpu} and \glspl{gpu},
greater availability of large datasets,
and developments in optimization algorithms for neural networks.
}
\Glspl{cnn} have become the state of the art
for many medical imaging tasks including
segmentation~\citep{cicek_3d_2016},
classification~\citep{lu_multimodal_2018},
reconstruction~\citep{chen_variable-density_2018}
and registration~\citep{shan_unsupervised_2018}.
Many of the network architectures and techniques have been adopted
from computer vision.

Compared to 2D \gls{rgb} images typically used in computer vision,
processing of medical images
such as \gls{mri}, \gls{us} or \gls{ct} presents different challenges.
These include a lack of labels for large datasets,
high computational costs (as the data is typically volumetric),
and the use of metadata to describe the physical size and
position of voxels.

Open-source frameworks for training \glspl{cnn} with medical images
have been built on top of
TensorFlow~\citep{abadi_tensorflow_2016,pawlowski_dltk_2017,gibson_niftynet_2018}.
Recently, the popularity of PyTorch~\citep{paszke_pytorch_2019} has
increased among researchers due to its improved usability compared to
TensorFlow~\citep{he_state_2019}, driving the need for open-source
tools compatible with PyTorch.
To reduce duplication of effort among research groups,
improve experimental reproducibility
and encourage open-science practices,
we have developed TorchIO:
an open-source Python library for efficient loading,
preprocessing, augmentation, and patch-based sampling of medical images
designed to be integrated into deep learning workflows.

TorchIO is a compact and modular library that can be seamlessly used alongside
higher-level deep learning frameworks for medical imaging, such as \gls{monai}.
It removes the need for researchers to code their own preprocessing
pipelines from scratch, which might be error-prone due to the complexity of
medical image representations.
Instead, it allows researchers to focus on their experiments,
supporting experiment reproducibility and traceability of their work,
and standardization of the methods used to process medical images for deep learning.
    \subsection{Motivation}

The nature of medical images makes it difficult to rely on
a typical computer-vision pipeline for neural network training.
In \cref{sec:challenges}, we describe challenges related to
medical images that need to be overcome when designing deep learning workflows.
In \cref{sec:frameworks}, we justify the choice of PyTorch as the main deep
learning framework dependency of TorchIO.

\subsubsection{Challenges in medical image processing for deep learning}
\label{sec:challenges}

In practice, multiple challenges must be addressed
when developing deep learning algorithms for medical images:
1) handling metadata related to physical position and size,
2) lack of large labeled datasets,
3) high computational costs due to data multidimensionality and
4) lack of consensus for best normalization practices.
%
\textcolor{rev1}{These challenges are very common in medical imaging and
require certain features that may not be implemented
in more general-purpose image processing frameworks such as
Albumentations~\citep{buslaev_albumentations_2020}
or TorchVision~\cite{paszke_pytorch_2019}.}

\paragraph{Metadata}
\label{sec:metadata}

In computer vision, picture elements, or \textit{pixels},
which are assumed to be square,
have a spatial relationship that comprises proximity and depth
according to both the arrangement of objects in the scene and camera placement.
In comparison, medical images are reconstructed such that the location of volume
elements, or cuboid-shaped \textit{voxels}, encodes a meaningful 3D spatial
relationship.
In simple terms, for 2D natural images, pixel vicinity does not necessarily
indicate spatial correspondence, while for medical images
spatial correspondence between nearby voxels can often be assumed.

Metadata, which encodes the physical size, spacing, and orientation of voxels,
determines spatial relationships between voxels~\citep{larobina_medical_2014}.
This information can provide meaningful context when performing medical image
processing, and is often implicitly or explicitly used in medical imaging
software.
Furthermore, metadata is often used to determine correspondence between images
as well as voxels within an image.
For example, registration algorithms for medical images typically work with
physical coordinates rather than voxel indices.

\begin{figure}
    \centering

    \begin{subfigure}{0.24\textwidth}
        \includegraphics[width=\linewidth]{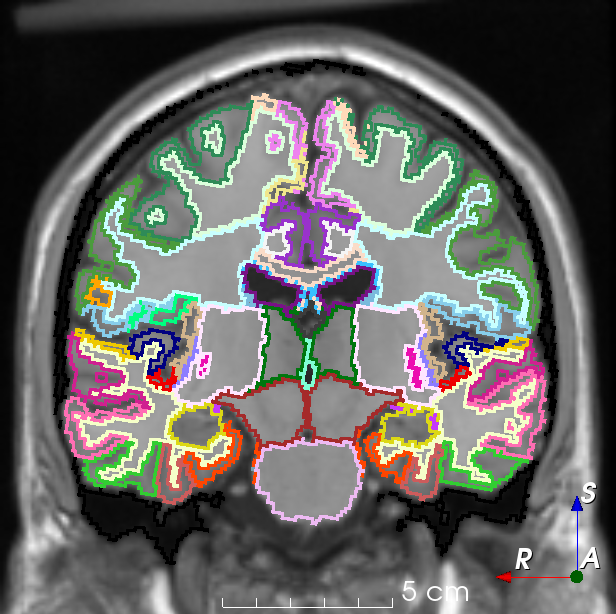}
        \caption{}
        \label{fig:meta_ok}
    \end{subfigure}
    \hfill
    \begin{subfigure}{0.24\textwidth}
        \includegraphics[width=\linewidth]{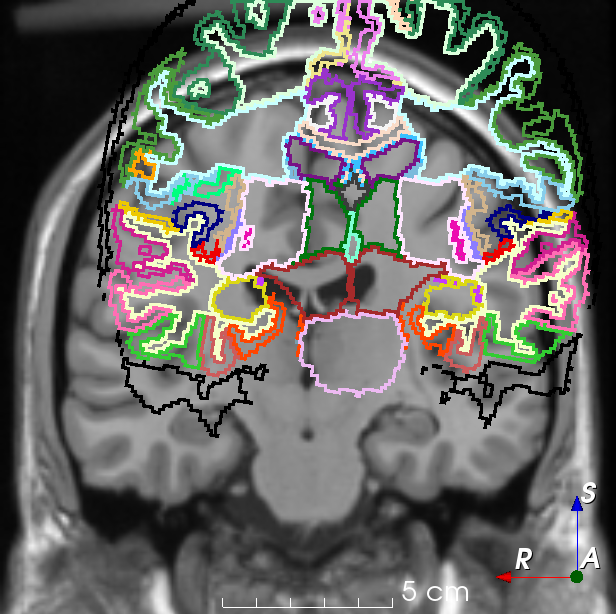}
        \caption{}
        \label{fig:meta_origin}
    \end{subfigure}
    \hfill
    \begin{subfigure}{0.24\textwidth}
        \includegraphics[width=\linewidth]{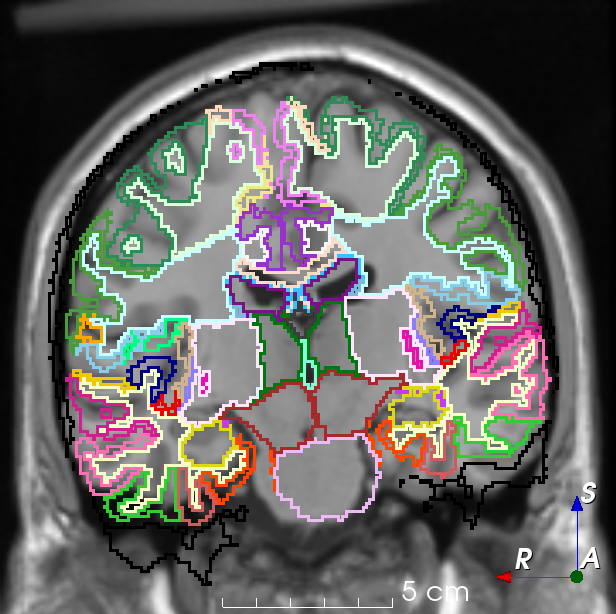}
        \caption{}
        \label{fig:meta_orientation}
    \end{subfigure}
    \hfill
    \begin{subfigure}{0.24\textwidth}
        \includegraphics[width=\linewidth]{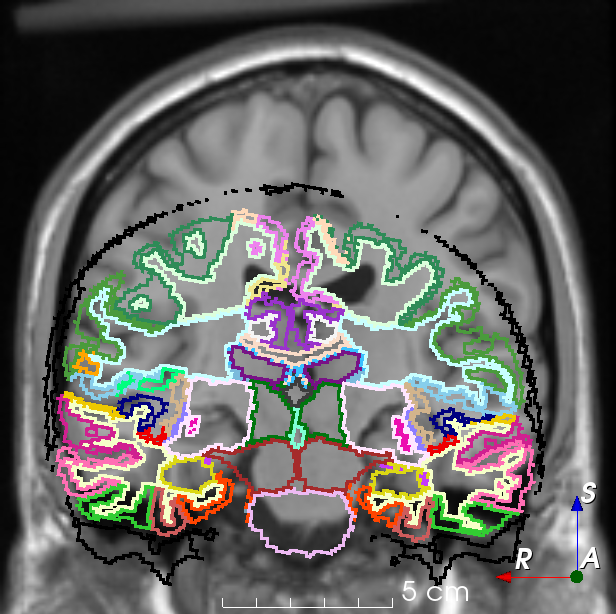}
        \caption{}
        \label{fig:meta_spacing}
    \end{subfigure}

    \caption{%
        Demonstration of the importance of spatial metadata in medical image
        processing.
        The size of both the \gls{mri} and the segmentation is $181 \times 181$.
        When spatial metadata is taken into account (a), images are correctly
        superimposed (only the borders of each region are shown for clarity
        purposes).
        Images are incorrectly superimposed if (b) origin,
        (c) orientation or (d) spacing are ignored.
    }
    \label{fig:metadata}
\end{figure}

\cref{fig:metadata} shows the superposition of an \gls{mri} and a corresponding
brain parcellation~\citep{cardoso_geodesic_2015} with the same
size ($181 \times 181$) but different origin, spacing and orientation.
A naïve user would assume that, given that the superimposition looks correct and
both images have the same size, they are ready for training.
However, the visualization is correct only because
3D Slicer~\citep{fedorov_3d_2012}, the software used for visualization,
is aware of the spatial metadata of the images.
As \glspl{cnn} generally do not take spatial metadata into account, training
using these images without preprocessing would lead to poor results.

\comment{
Medical images have an \textit{offset},
which is the position in mm of the first voxel in the file
with respect to some origin in the physical space.
For example, in \gls{mri}, the \textit{world} origin is usually set at the
magnet isocenter.
The \textit{spacing} is the distance between the voxels centroids along each
dimension.
Lastly, medical images have an \textit{orientation} to define the angles
between image axes and the spatial reference system.
These three attributes are required to interpret the spatial properties of the
images properly.
Typically, they are encoded in a 6-DOF\footnote{
    CT images acquired with a gantry tilt might require a 9-DOF matrix to
    additionally encode shear parameters~\cite{bibb_medical_2015}.
}, $4 \times 4$~affine matrix representing a Euclidean transform
that can be used to map voxel indices to physical
positions using homogeneous coordinates.
A medical image can then generally be defined by a 3D or 4D tensor
containing voxel data and a 2D matrix representing the spatial information.
The fourth dimension is commonly used
for one-hot encoding of labels,
for the results of applying different gradients in \gls{dmri}
or to encode time in video-\gls{us} or \gls{fmri}.
}

Medical images are typically stored in specialized formats such as
\gls{dicom} or
\gls{nifti}~\citep{larobina_medical_2014},
and commonly read and processed by medical imaging frameworks
such as SimpleITK~\citep{lowekamp_design_2013}
or NiBabel~\citep{brett_nipynibabel_2020}.

\paragraph{Limited training data}
Deep learning methods typically require large amounts of annotated data,
which are often scarce in clinical scenarios due to
concerns over patient privacy,
the financial and time burden associated with collecting data as part of a
clinical trial,
and the need for annotations from highly-trained and experienced raters.
Data augmentation techniques can be used to increase the size of
the training dataset artificially by applying different transformations to each
training instance while preserving the relationship to annotations.

Data augmentation performed in computer vision typically aims to simulate
variations in camera properties, \gls{fov}, or perspective.
Traditional data augmentation operations applied in computer vision include
geometrical transforms such as random rotation or zoom,
color-space transforms such as random channel swapping
or kernel filtering such as random Gaussian blurring.
Data augmentation is usually performed on the fly, i.e., every time an image is
loaded from disk during training.

Several computer vision libraries supporting data augmentation
have appeared recently, such as
Albumentations~\citep{buslaev_albumentations_2020},
or \texttt{imgaug}~\citep{jung_imgaug_2020}.
PyTorch also includes
some computer vision transforms,
mostly implemented as Pillow wrappers~\citep{wiredfool_pillow_2016}.
However, none of these libraries support reading or transformations for 3D images.
Furthermore, medical images are almost always grayscale,
therefore color-space transforms are not applicable.
Additionally, cropping and scaling are more challenging to apply to medical images
without affecting the spatial relationships of the data.
Metadata should usually be considered when applying these transformations to medical images.

In medical imaging, the purpose of data augmentation is designed to simulate anatomical
variations and scanner artifacts.
Anatomical variation and sample position can be simulated using spatial transforms such as elastic
deformation, lateral flipping, or affine transformations.
Some artifacts are unique to specific medical image modalities.
For example,
ghosting artifacts will be present in \gls{mri}
if the patient moves during acquisition, and metallic implants often produce
streak artifacts in \gls{ct}.
Simulation of these artifacts can be useful when performing augmentation
on medical images.

\paragraph{Computational costs}
\label{sec:computation}
The number of pixels in 2D images used in deep learning
is rarely larger than one million.
For example, the input size of several popular image classification
models is $224 \times 224 \times 3 = \num{150528}$ pixels
(\SI{588}{\kibi\byte} if 32 bits per pixel are used).
In contrast, 3D medical images often contain hundreds of
millions of voxels, and downsampling might not be acceptable when small details
should be preserved.
For example, the size of a high-resolution lung \gls{ct}-scan
used for quantifying \acrlong{copd} damage in a research setting,
with spacing $0.66 \times 0.66 \times 0.30$ mm,
is $512 \times 512 \times 1069 = \num{280231936}$ voxels
(\SI{1.04}{\gibi\byte} if 32 bits per voxel are used).

In computer vision applications, images used for training are grouped in
batches whose size is often in the order of
hundreds~\citep{krizhevsky_imagenet_2012}
or even thousands~\citep{chen_simple_2020} of training instances,
depending on the available \gls{gpu} memory.
In medical image applications, batches rarely contain more than
one~\citep{cicek_3d_2016} or two~\citep{milletari_v-net_2016} training instances
due to their larger memory footprint compared to natural images.
This reduces the utility of techniques
such as batch normalization,
which rely on batches being large enough to estimate
dataset variance appropriately~\citep{ioffe_batch_2015}.
Moreover, large image size and small batches result in longer training time,
hindering the experimental cycle that is necessary for hyperparameter
optimization.
In cases where \gls{gpu} memory is limited and the network architecture is large,
it is possible that not even the entirety of a single volume can be processed during
a training iteration.
To overcome this challenge, it is common in medical imaging to train using
subsets of the image, or image \textit{patches},
randomly extracted from the volumes.

Networks can be trained with 2D slices extracted from 3D volumes,
aggregating the inference results to generate a 3D volume~\citep{lucena_convolutional_2019}.
This can be seen as a specific case of patch-based training,
where the size of the patches along a dimension is one.
Other methods extract volumetric patches for training,
that are often cubes,
if the voxel spacing is isotropic~\citep{li_compactness_2017},
or cuboids adapted to the
anisotropic spacing of the training images~\citep{nikolov_deep_2018}.
\comment{  
Recent techniques such as
gradient checkpointing~\citep{chen_training_2016},
automatic mixed precision~\citep{micikevicius_mixed_2018}
or reversible layers~\citep{brugger_partially_2019}
can reduce the memory burden when training with large 3D images.
}

\paragraph{Transfer learning and normalization}

One can pre-train a network on a large dataset of natural images such as
ImageNet~\citep{deng_imagenet_2009},
which contains more than 14 million labeled images,
and fine-tune on a custom, much smaller target dataset.
This is a typical use of transfer learning
in computer vision~\cite{weiss_survey_2016}.
The literature has reported mixed results using transfer learning to apply models pretrained on
natural images to medical images~\citep{cheplygina_cats_2019,raghu_transfusion_2019}.

%
In computer vision, best practice is to normalize each training instance before
training, using statistics computed from the whole training
dataset~\cite{krizhevsky_imagenet_2012}.
Preprocessing of medical images is often performed on a per-image basis,
and best practice is to take into account the bimodal nature of medical images
(i.e., that an image has a background and a foreground).

Medical image voxel intensity values can be encoded with different data types
and intensity ranges, and the meaning of a specific value can vary
between different modalities, sequence acquisitions, or scanners.
Therefore, intensity normalization methods for medical images often involve more complex parameterization of intensities than those used for natural images~\citep{nyul_standardizing_1999}.

\subsubsection{Deep learning frameworks}
\label{sec:frameworks}

There are currently two major generic deep learning frameworks:
TensorFlow~\citep{abadi_tensorflow_2016}
and PyTorch~\citep{paszke_pytorch_2019},
primarily maintained by Google and Facebook, respectively.
Although TensorFlow has traditionally been the primary choice for both research
and industry, PyTorch has recently seen a substantial increase in popularity,
especially among the research community~\citep{he_state_2019}.

PyTorch is often preferred by the research community as it is \textit{pythonic},
i.e., its design, usage, and \gls{api} follow the conventions of plain Python.
Moreover, the \gls{api} for tensor operations follows a similar paradigm to the
one for NumPy multidimensional arrays, which is the primary array programming
library for the Python language~\citep{van_der_walt_numpy_2011}.
In contrast, for TensorFlow, researchers need to become familiar with new design
elements such as sessions, placeholders, feed dictionaries,
gradient tapes and static graphs.
In PyTorch, objects are standard Python classes and variables, and a dynamic
graph makes debugging intuitive and familiar to anyone already using Python.
These differences have decreased with the recent release of TensorFlow 2,
whose eager mode makes usage reminiscent of Python.

TorchIO was designed to be in the style of PyTorch and uses several
of its tools to reduce the barrier to learning how to
use TorchIO for those researchers already familiar with PyTorch.

    \subsection{Related work}

NiftyNet~\citep{gibson_niftynet_2018} and the \gls{dltk}~\citep{pawlowski_dltk_2017}
are deep learning frameworks
designed explicitly for medical image processing using the TensorFlow~1
platform.
Both of them are no longer being actively maintained.
They provide implementations of some popular network architectures
such as U-Net~\citep{cicek_3d_2016}, and can be used to train
3D \glspl{cnn} for different tasks.
\textcolor{rev2}{For example, NiftyNet was used to train a 3D residual network
for brain parcellation~\citep{li_compactness_2017}, and \gls{dltk} was used to
perform multi-organ segmentation on \gls{ct} and
\gls{mri}~\citep{valindria_multi-modal_2018}.}
\comment{
NiftyNet includes some preprocessing and augmentation operations
specifically for medical images, such as
histogram standardization~\citep{nyul_new_2000}
and random bias field
augmentation~\citep{van_leemput_automated_1999,sudre_longitudinal_2017},
both implemented using NumPy.
It is designed to be used through a high-level configuration file,
making its usage slightly cumbersome for researchers
who desire lower-level access to alter or augment its features,
specific network architectures or individual components.
The last set of substantial commits on the \gls{dltk} repository
are from June 2018, which suggests that the code is not actively maintained.
Development of NiftyNet is officially discontinued.
}

The \texttt{medicaltorch}
library~\citep{christian_s_perone_peronemedicaltorch_2018}
closely follows the PyTorch design, and provides some functionalities for
preprocessing, augmentation and training of medical images.
However, it does not leverage the power of specialized medical image processing
libraries, such as SimpleITK~\citep{lowekamp_design_2013}, to process volumetric images.
\comment{
For example, random 3D rotation is performed to the volume
slice by slice along a specified axis using PyTorch.
If rotations around more axes are desired, computation time will
increase linearly with the number of rotations.
Moreover, interpolation applied at multiple resampling operations degrades
the image.
In contrast, multiple rotations, translations, shearings, etc. may be
composed into a single affine transform and applied once in 3D,
reducing resampling artifacts and computational costs.
}
Similar to \gls{dltk}, this library has not seen much activity since 2018.

The \texttt{batchgenerators}
library~\citep{isensee_batchgenerators_2020},
\textcolor{rev2}{used within the popular medical segmentation framework
nn-UNet~\citep{isensee_nnu-net_2021},}
includes custom dataset and data loader classes for
multithreaded loading of 3D medical images,
implemented before data loaders were available in PyTorch.
In the usage examples from GitHub,
preprocessing is applied to the whole dataset before training.
Then, spatial data augmentation is performed at the volume level,
from which one patch is extracted
and intensity augmentation is performed at the patch level.
In this approach, only one patch is extracted per volume,
diminishing the efficiency of training pipelines.
Transforms in \texttt{batchgenerators} are mostly implemented using
NumPy~\citep{van_der_walt_numpy_2011} and
SciPy~\citep{virtanen_scipy_2020}.

\textcolor{rev2}{%
More recently, a few PyTorch-based libraries for deep learning and medical images
have appeared.
There are two other libraries, developed in parallel to TorchIO,
focused on data preprocessing and augmentation.
Rising\fnurl{https://github.com/PhoenixDL/rising} is a library for data
augmentation entirely written in PyTorch,
which allows for gradients to be propagated through the transformations and
perform all computations on the \gls{gpu}.
%
However, this means specialized medical imaging libraries
such as SimpleITK cannot be used.
\texttt{pymia}~\citep{jungo_pymia_2021} provides features
for data handling (loading, preprocessing, sampling) and evaluation.
It is compatible with TorchIO transforms,
which are typically leveraged for data augmentation,
as their data handling is more focused on preprocessing.
\texttt{pymia} can be easily integrated into either
PyTorch or TensorFlow pipelines.
It was recently used to assess
the suitability of evaluation metrics for medical image
segmentation~\citep{kofler_are_2021}.
}

\gls{monai}~\citep{nic_ma_project-monaimonai_2021}
and Eisen~\citep{mancolo_eisen_2020}
are PyTorch-based frameworks for deep learning workflows with medical images.
Similar to NiftyNet and \gls{dltk}, they include implementation of network
architectures, transforms, and higher-level features to perform training and
inference.
\textcolor{rev2}{For example, \gls{monai} was recently used for brain
segmentation on fetal \gls{mri}~\citep{ranzini_monaifbs_2021}.
As these packages are solving a large problem, i.e., that of workflow
in deep learning for medical images, they do not contain all of the
data augmentation transforms present in TorchIO.
However, it is important to note that an end user does not need to
select only one open-source package, as
}
TorchIO transforms are compatible with both Eisen and \gls{monai}.

\textcolor{rev2}{%
TorchIO is a library that specializes in preprocessing
and augmentation using PyTorch, focusing on ease of use for researchers.
This is achieved by providing
a PyTorch-like \gls{api},
comprehensive documentation with many usage examples,
and tutorials showcasing different features,
and by actively addressing feature requests and bug reports from
the many users that have already adopted TorchIO.
This is in contrast with other modern libraries released after TorchIO
such as \gls{monai}, which aims to deliver a larger umbrella of
functionalities including federated learning or active learning, but may have
slower development and deployment.
}

    \section{Methods}

We developed TorchIO, a Python library that focuses on data loading and
augmentation of medical images in the context of deep learning.

TorchIO is a unified library to load and augment data that
makes explicit use of medical image properties,
and is flexible enough to be used for different loading workflows.
It can accelerate research by avoiding the need to code a processing pipeline
for medical images from scratch.

In contrast with Eisen or \gls{monai},
we do not implement network architectures, loss functions or training workflows.
This is to limit the scope of the library and to enforce modularity between
training of neural networks and preprocessing and data augmentation.

Following the PyTorch philosophy~\citep{paszke_pytorch_2019},
we designed TorchIO with an emphasis on simplicity and usability
while reusing PyTorch classes and infrastructure where possible.
Note that, although we designed TorchIO following PyTorch style,
the library could also be used with other deep learning platforms such as
TensorFlow or Keras~\citep{chollet_keras_2015}.

TorchIO makes use of open-source medical imaging software platforms.
Packages were selected to reduce the number of required external
dependencies and the need to re-implement basic medical imaging
processing operations (image loading, resampling, etc.).

\begin{figure}
    \centering
    \includegraphics[width=\linewidth]{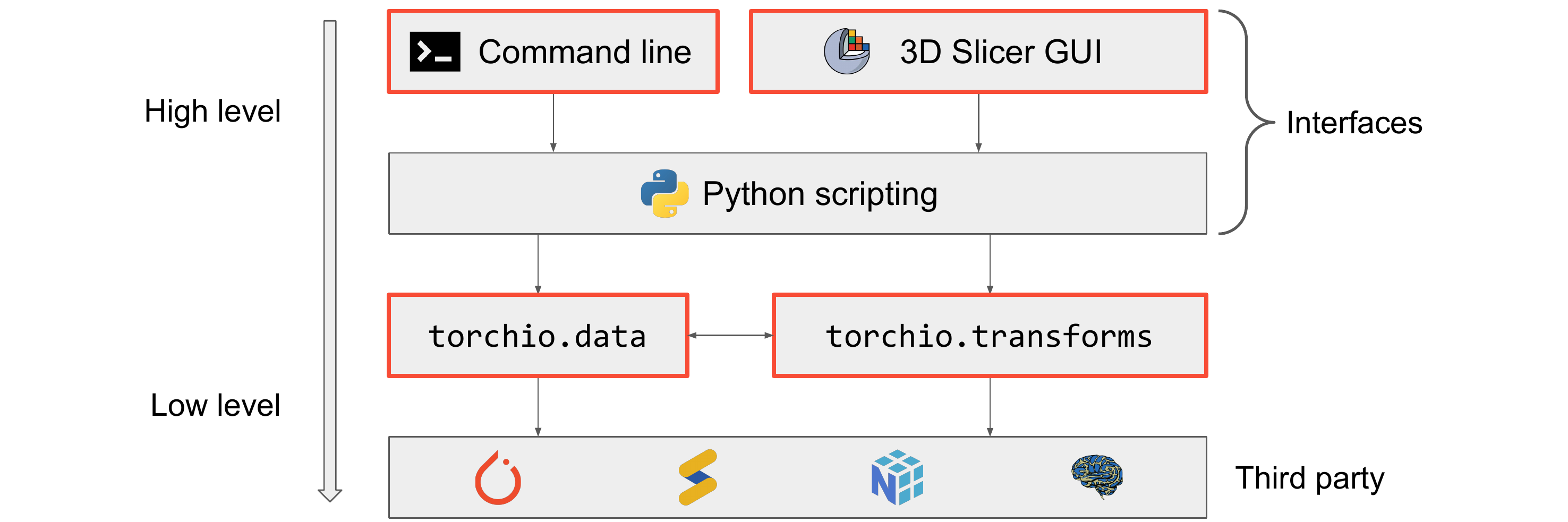}
    \caption{%
        General diagram of TorchIO, its dependencies and its interfaces.
        Boxes with a red border
        (\protect\tikz[baseline=-0.5ex]\protect\draw[thick, color = pytorch_orange] (0,0) -- (0.5,0);)
        represent elements implemented in TorchIO.
        Logos indicate lower-level Python libraries used by TorchIO.
        \protect\logo{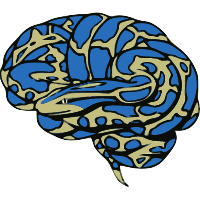}:~NiBabel\citep{brett_nipynibabel_2020};
        \protect\logo{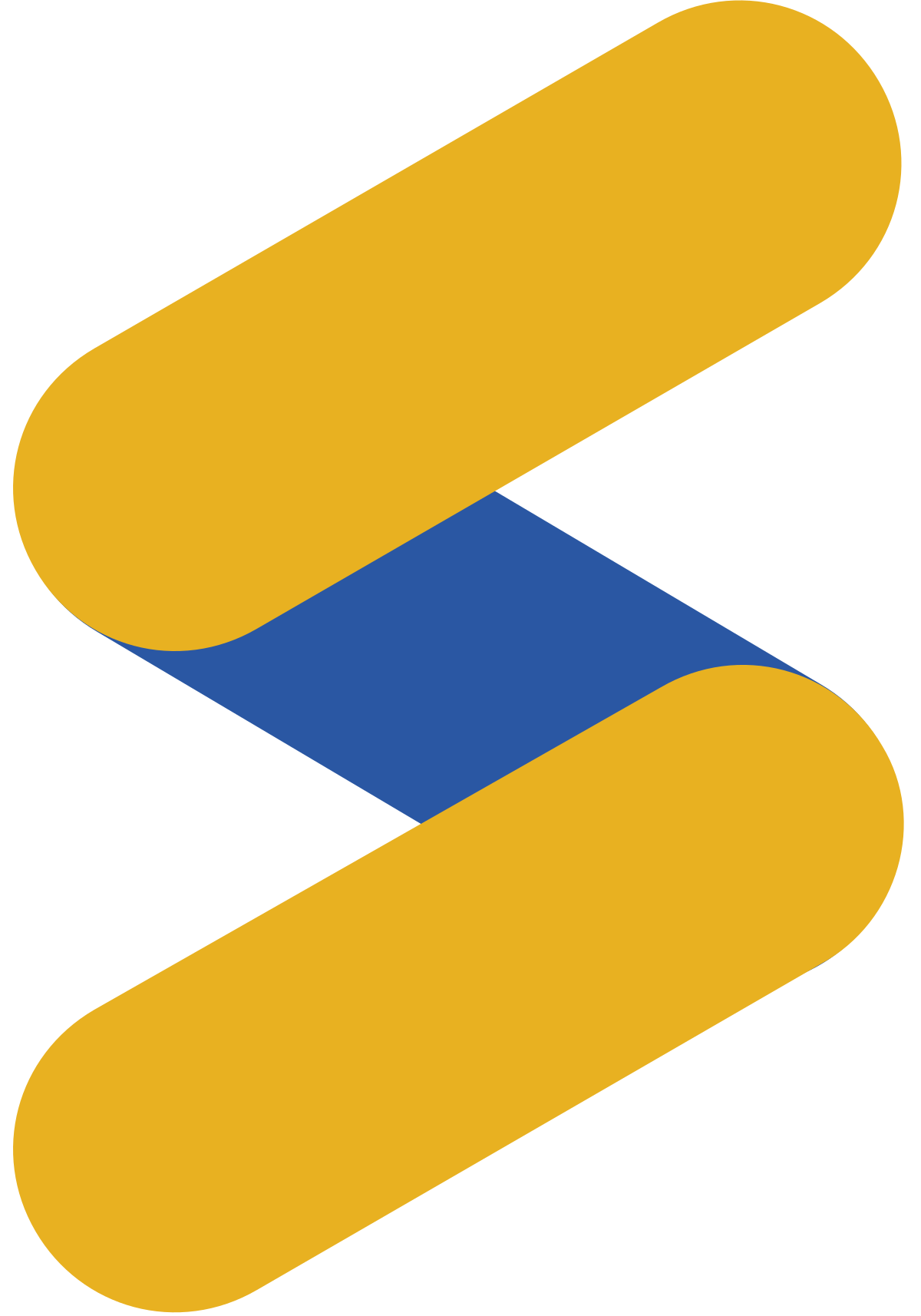}:~SimpleITK\citep{lowekamp_design_2013};
        \protect\logo{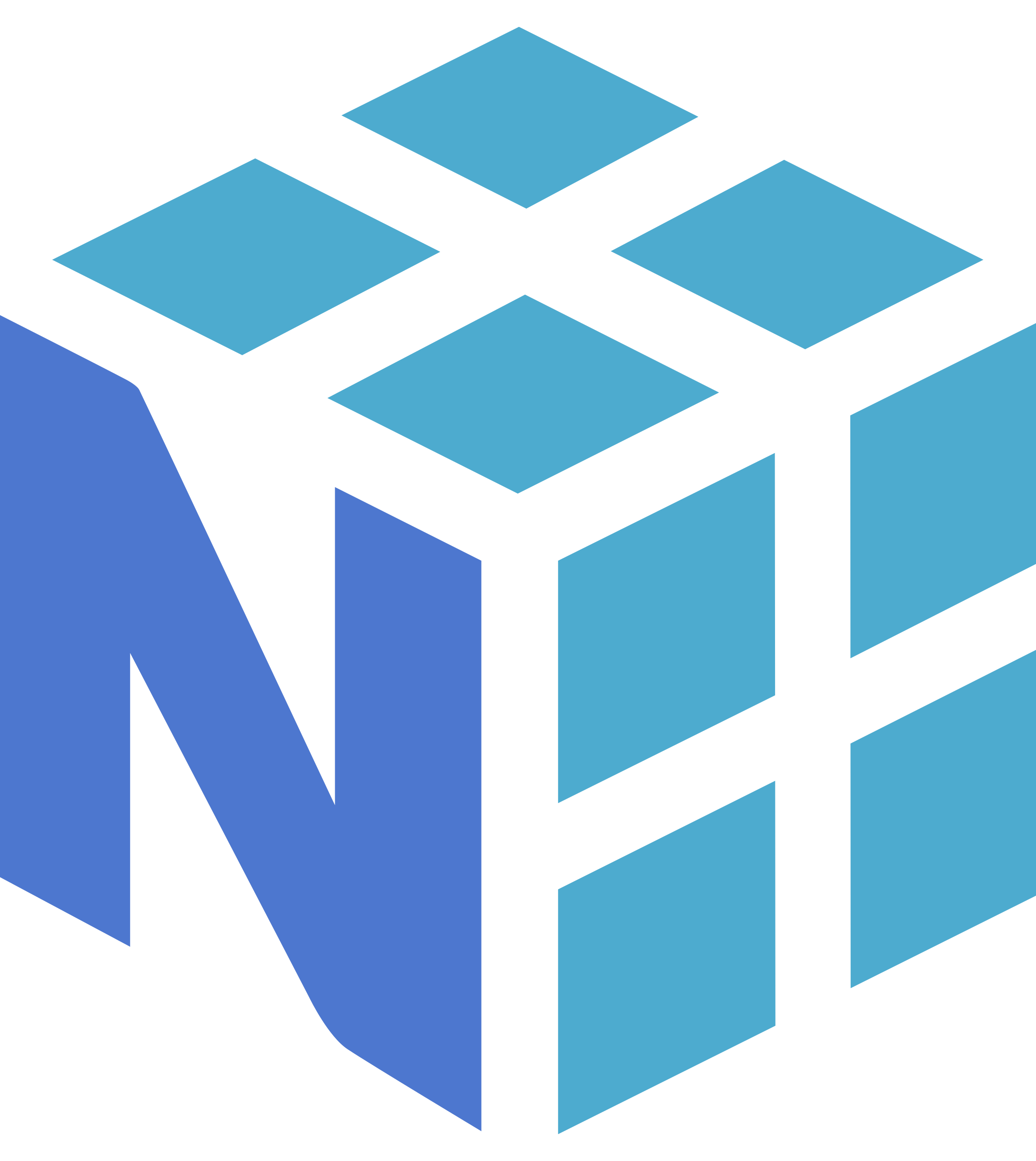}:~NumPy\citep{van_der_walt_numpy_2011};
        \protect\logo{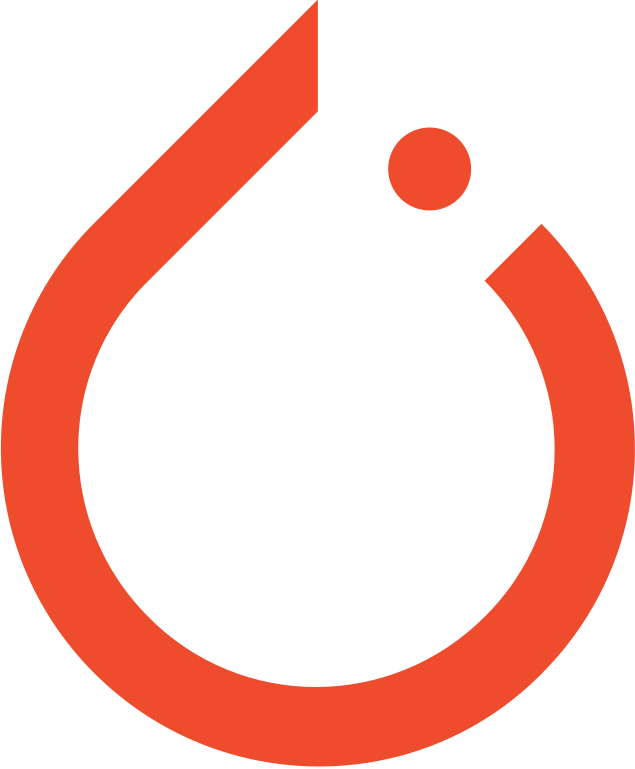}:~PyTorch\citep{paszke_pytorch_2019}.
    }
    \label{fig:torchio}
\end{figure}

TorchIO features are divided into two categories: data structures and
input/output (\texttt{torchio.data}),
and transforms for preprocessing and augmentation (\texttt{torchio.transforms}).
\cref{fig:torchio} represents a diagram of the codebase and the different
interfaces to the library.

\subsection{Data}

\subsubsection{Input/Output}

TorchIO uses the medical imaging libraries NiBabel and SimpleITK to read and
write images.
Dependency on both is necessary to ensure broad support of image formats.
For instance, NiBabel does not support reading \gls{png} files,
while SimpleITK does not support some neuroimaging-specific formats.

TorchIO supports up to 4D images, i.e., 2D or 3D single-channel or multi-channel
data such as X-rays, \gls{rgb} histological slides, microscopy stacks,
multispectral images, \gls{ct}-scans \gls{fmri} and \gls{dmri}.

\subsubsection{Data structures}
\label{sec:data_structures}

\paragraph{Image}

The \texttt{Image} class, representing one medical image,
stores a 4D tensor, whose voxels encode, e.g., signal intensity or segmentation
labels, and the corresponding affine transform,
typically a rigid (Euclidean) transform, to convert
voxel indices to world coordinates in millimeters.
Arbitrary fields such as acquisition parameters may also be stored.

Subclasses are used to indicate specific types of images,
such as \texttt{ScalarImage} and \texttt{LabelMap},
which are used to store, e.g., \gls{ct} scans and segmentations, respectively.

An instance of \texttt{Image} can be created using a filepath, a PyTorch tensor,
or a NumPy array.
This class uses lazy loading, i.e., the data is not loaded from disk at
instantiation time.
Instead, the data is only loaded when needed for an operation
(e.g., if a transform is applied to the image).

\begin{figure}
    \centering
    \includegraphics[%
        width=\linewidth,
        trim = {0 0 4.8cm 0},
        clip,
    ]{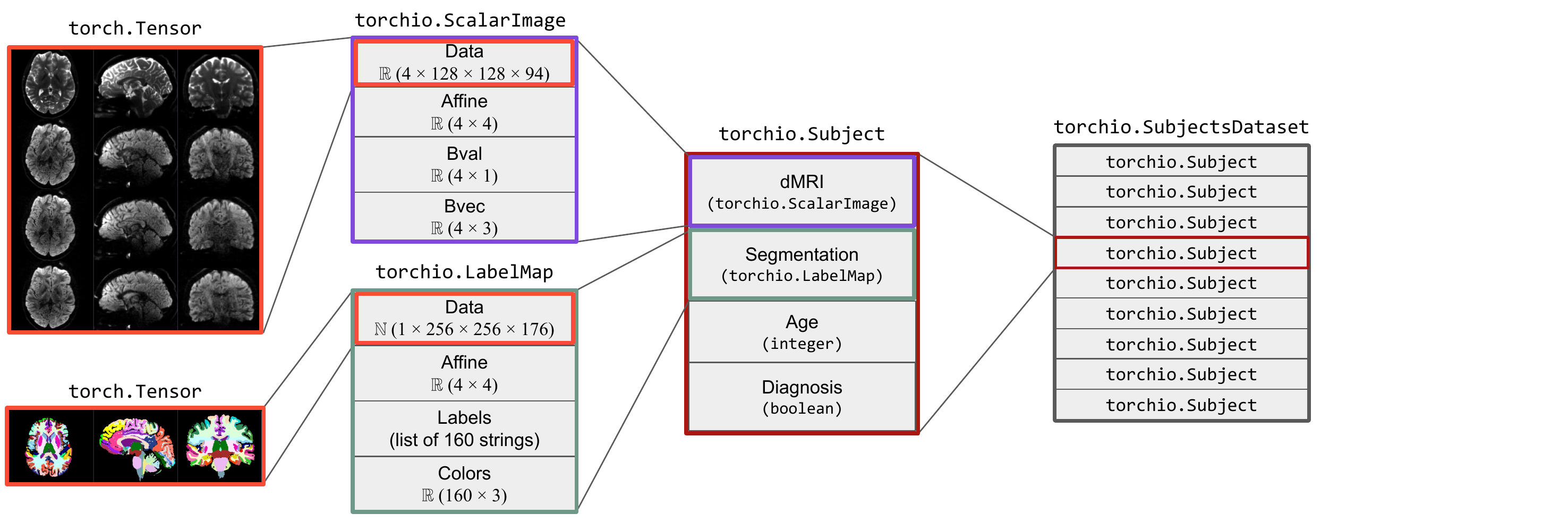}
    \caption{%
        Usage example of \texttt{ScalarImage}, \texttt{LabelMap},
        \texttt{Subject} and \texttt{SubjectsDataset}.
        The images store a 4D \gls{dmri} and a brain parcellation,
        and other related metadata.
    }
    \label{fig:data_structures}
\end{figure}

\cref{fig:data_structures} shows two instances of \texttt{Image}.
The instance of \texttt{ScalarImage} contains a 4D tensor representing a
\gls{dmri}, which contains four 3D volumes (one per gradient direction),
and the associated affine matrix.
Additionally, it stores the strength and direction for each of the four
gradients.
The instance of \texttt{LabelMap} contains a brain parcellation of the same
subject, the associated affine matrix, and the name and color of each brain
structure.

\paragraph{Subject}

The \texttt{Subject} class stores instances of \texttt{Image} associated to
a subject, e.g., a human or a mouse.
As in the \texttt{Image} class, \texttt{Subject} can store arbitrary fields
such as age, diagnosis or ethnicity.

\paragraph{Subjects dataset}

The \texttt{SubjectsDataset} inherits from the PyTorch \texttt{Dataset}.
It contains the list of subjects and optionally a transform
to be applied to each subject after loading.
When \texttt{SubjectsDataset} is queried for a specific subject,
the corresponding set of images are loaded,
a transform is applied to the images
and the instance of \texttt{Subject} is returned.

\begin{figure}[ht]
    \centering

    \begin{subfigure}{\textwidth}
        \includegraphics[%
            width=0.98\linewidth,
            trim = {0 1cm 0 1cm},
            clip
        ]
        {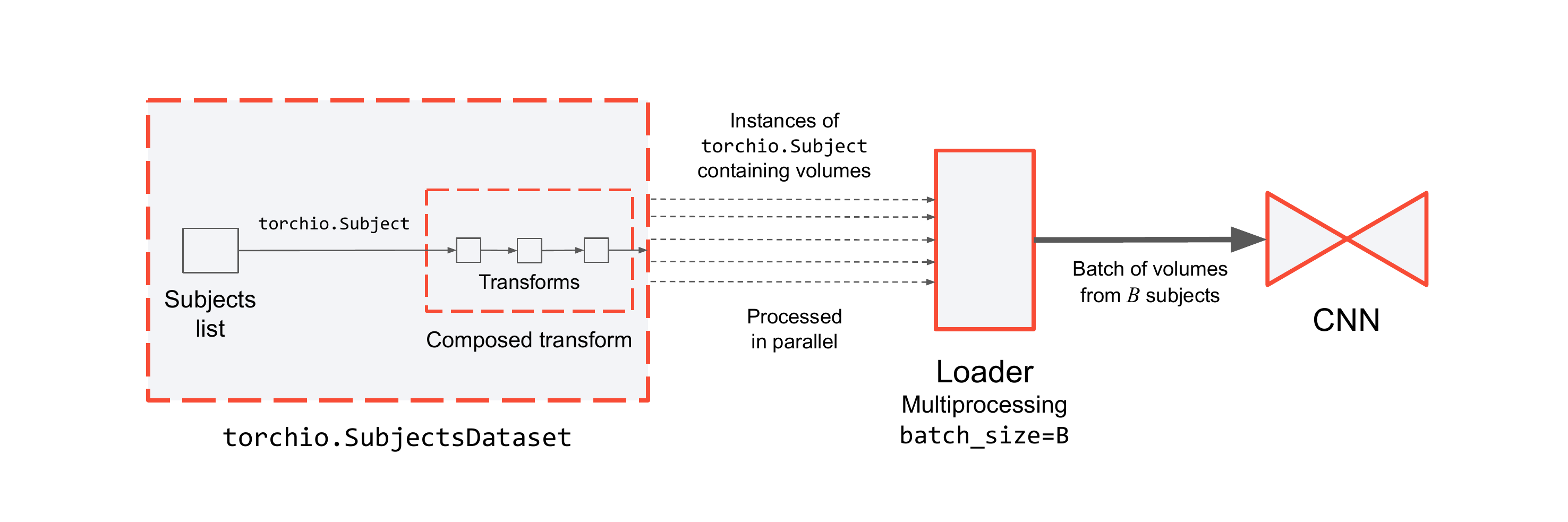}
        \caption{Training with whole volumes}
        \label{fig:diagram_volumes}
    \end{subfigure}

    \begin{subfigure}{\textwidth}
        \includegraphics[%
            width=0.98\linewidth,
            trim = {0 0.5cm 0 0},
            clip
        ]
        {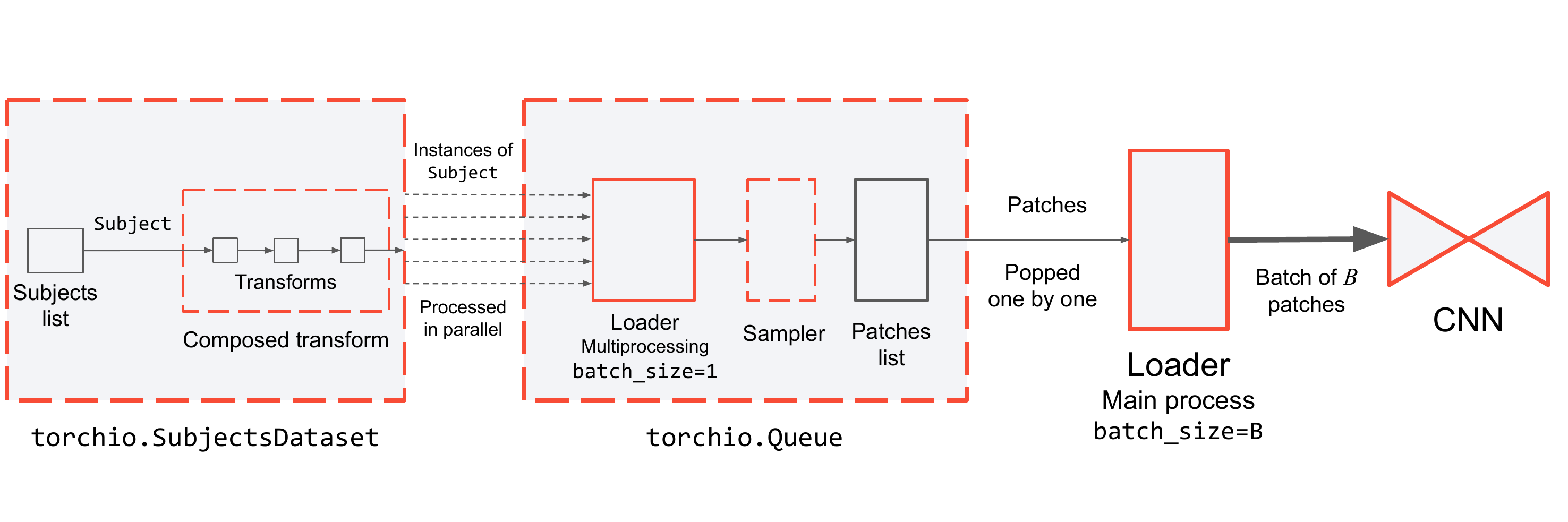}
        \caption{Training with patches}
        \label{fig:diagram_patches}
    \end{subfigure}

    \caption{%
        Diagram of data pipelines for training with whole volumes (top)
        and patches (bottom).
        Boxes with a red border represent PyTorch classes
        (\protect\tikz[baseline=-0.5ex]\protect\draw[thick, color = pytorch_orange] (0,0) -- (0.5,0);)
        or TorchIO classes that inherit from PyTorch classes
        (\protect\tikz[baseline=-0.5ex]\protect\draw[dashed, thick, color = pytorch_orange] (0,0) -- (0.5,0);).
        }
    \label{fig:diagram}
\end{figure}

For parallel loading, a PyTorch \texttt{DataLoader} may be used.
This loader spawns multiple processes, each of which contains a
shallow copy of the \texttt{SubjectsDataset}.
Each copy is queried for a different subject, therefore loading and
transforming is applied to different subjects in parallel on the \gls{cpu} (\cref{fig:diagram_volumes}).

An example of subclassing \texttt{SubjectsDataset} is
\texttt{torchio.datasets.IXI}, which may be used to download
the \gls{ixi} dataset\fnurl{https://brain-development.org/ixi-dataset/}.

\subsubsection{Patch-based training}
\label{sec:patches}

Memory limitations often require
training and inference steps to be performed using
image subvolumes or \textit{patches} instead of the whole volumes,
as explained in \cref{sec:computation}.
In this section, we describe how TorchIO implements patch-based training via
image sampling and queueing.

\paragraph{Samplers}

A sampler takes as input an instance of \texttt{Subject} and returns a version
of it whose images have a reduced \gls{fov}, i.e.,
the new images are subvolumes, also called windows or \textit{patches}.
For this, a \texttt{PatchSampler} may be used.

Different criteria may be used to select the center voxel of each output
patch.
A \texttt{UniformSampler} selects a voxel as the center at random with all
voxels having an equal probability of being selected.
A \texttt{WeightedSampler} selects the patch center according to a
probability distribution image defined over all voxels,
which is passed as input to the sampler.

At testing time, images are sampled such that a dense inference can be performed on the input volume.
A \texttt{GridSampler} can be used to sample patches such that the center voxel is selected using a set stride.
In this way, sampling over the entire volume is ensured.
The potentially-overlapping inferred patches can be passed to a \texttt{GridAggregator} that builds
the resulting volume patch by patch (or batch by batch).  

\paragraph{Queue}

A training iteration (i.e., forward and backward pass) performed on a \gls{gpu}
is usually faster than loading, preprocessing, augmenting, and cropping a volume
on a \gls{cpu}.
Most preprocessing operations could be performed using a \gls{gpu},
but these devices are typically reserved for training the \gls{cnn} so that
the batch size and input tensor can be as large as possible.
Therefore, it is beneficial to prepare
(i.e., load, preprocess and augment) the volumes using multiprocessing
\gls{cpu} techniques in parallel with the forward-backward passes of a training iteration.

Once a volume is appropriately prepared, it is computationally beneficial to
sample multiple patches from a volume rather than having to prepare the same
volume each time a patch needs to be extracted.
The sampled patches are then stored in a buffer or \textit{queue} until
the next training iteration, at which point they are loaded onto the \gls{gpu}
to perform an optimization iteration.
For this, TorchIO provides the \texttt{Queue} class, which inherits from the
PyTorch \texttt{Dataset} (\cref{fig:diagram_patches}).
In this queueing system,
samplers behave as generators that yield patches
from volumes contained in the \texttt{SubjectsDataset}.

The end of a training epoch is defined as the moment after which patches from
all subjects have been used for training.
At the beginning of each training epoch,
the subjects list in the \texttt{SubjectsDataset} is shuffled,
as is typically done in machine learning pipelines to increase variance
of training instances during model optimization.
%
A PyTorch loader begins by shallow-copying the dataset to each subprocess.
Each worker subprocess loads and applies image transforms to the volumes in parallel.
A patches list is filled with patches extracted by the sampler,
and the queue is shuffled once it has reached a specified maximum length so that batches are composed of patches from different
subjects.
The internal data loader continues querying the \texttt{SubjectsDataset} using multiprocessing.
The patches list, when emptied, is refilled with new patches.
A second data loader, external to the queue,
may be used to collate batches of patches stored in the queue,
which are passed to the neural network.

\subsection{Transforms}
\label{sec:transforms}

\begin{figure}
    \centering
    \captionsetup[subfigure]{aboveskip=0pt, belowskip=3pt}

    \begin{subfigure}{0.45\textwidth}
        \includegraphics[width=0.98\linewidth]{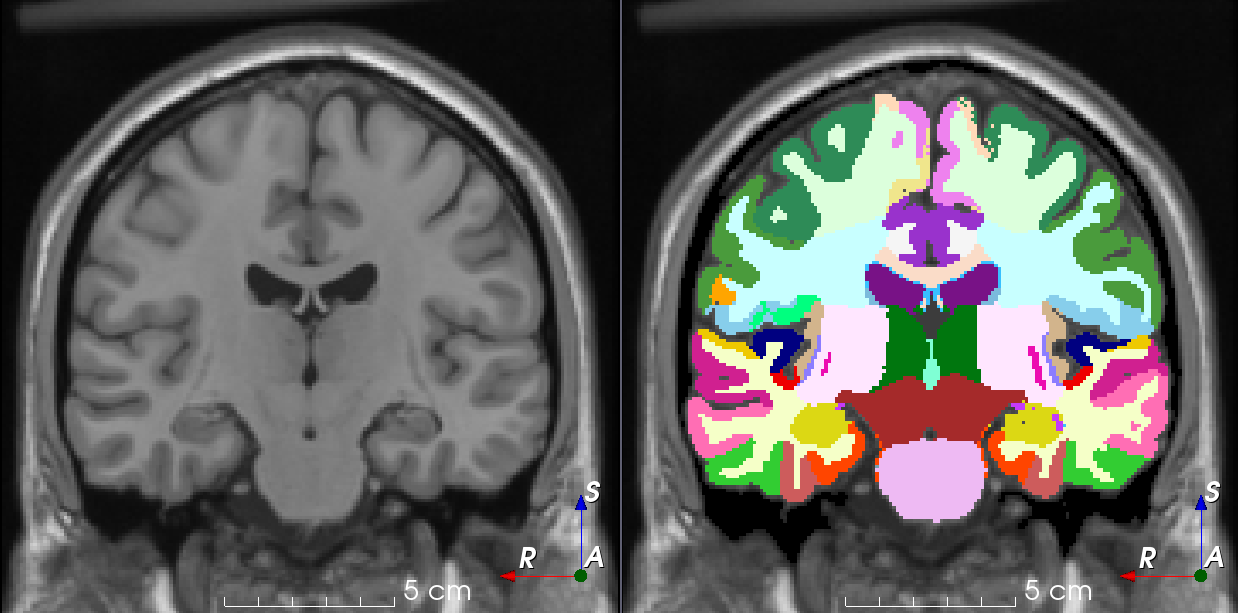}
        \caption{Original image and segmentation}
        \label{fig:original}
    \end{subfigure}%
    \hspace{2em}%
    \begin{subfigure}{0.45\textwidth}
        \includegraphics[width=0.98\linewidth]{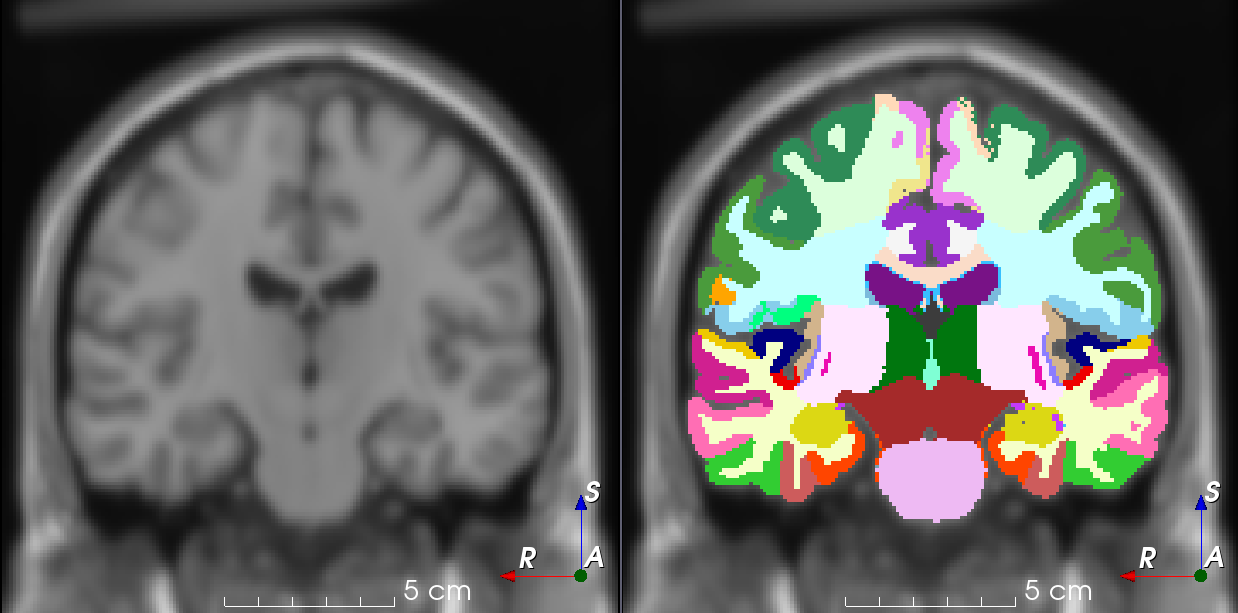}
        \caption{Random blur}
        \label{fig:rblur}
    \end{subfigure}

    \begin{subfigure}{0.45\textwidth}
        \includegraphics[width=0.98\linewidth]{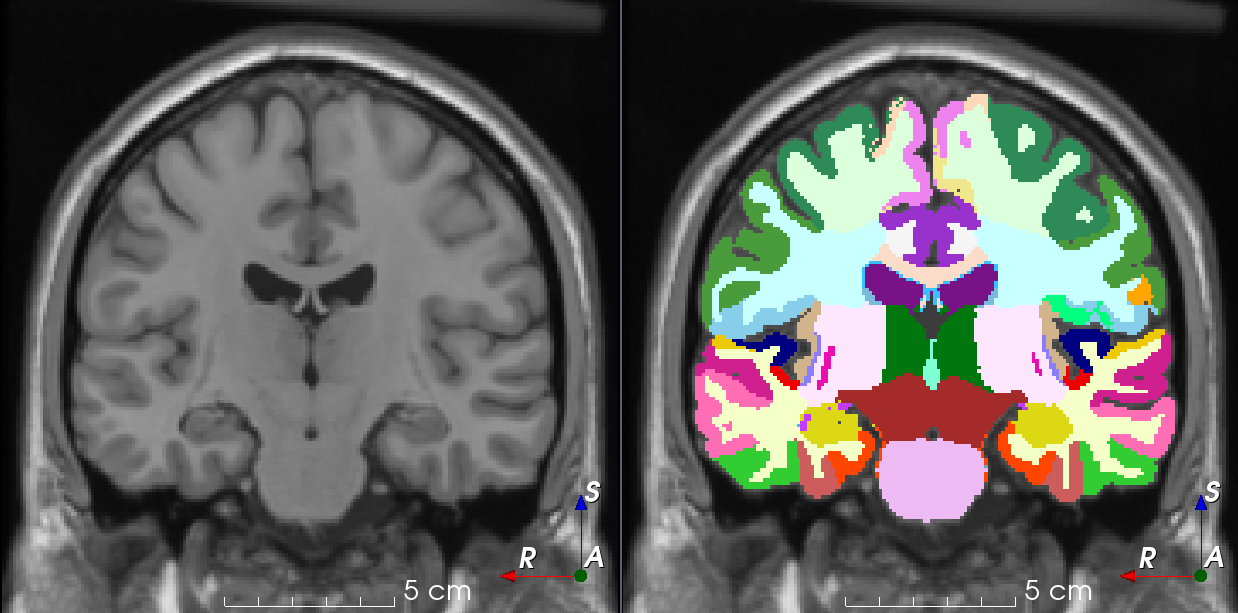}
        \caption{Random flip}
        \label{fig:rflip}
    \end{subfigure}%
    \hspace{2em}%
    \begin{subfigure}{0.45\textwidth}
        \includegraphics[width=0.98\linewidth]{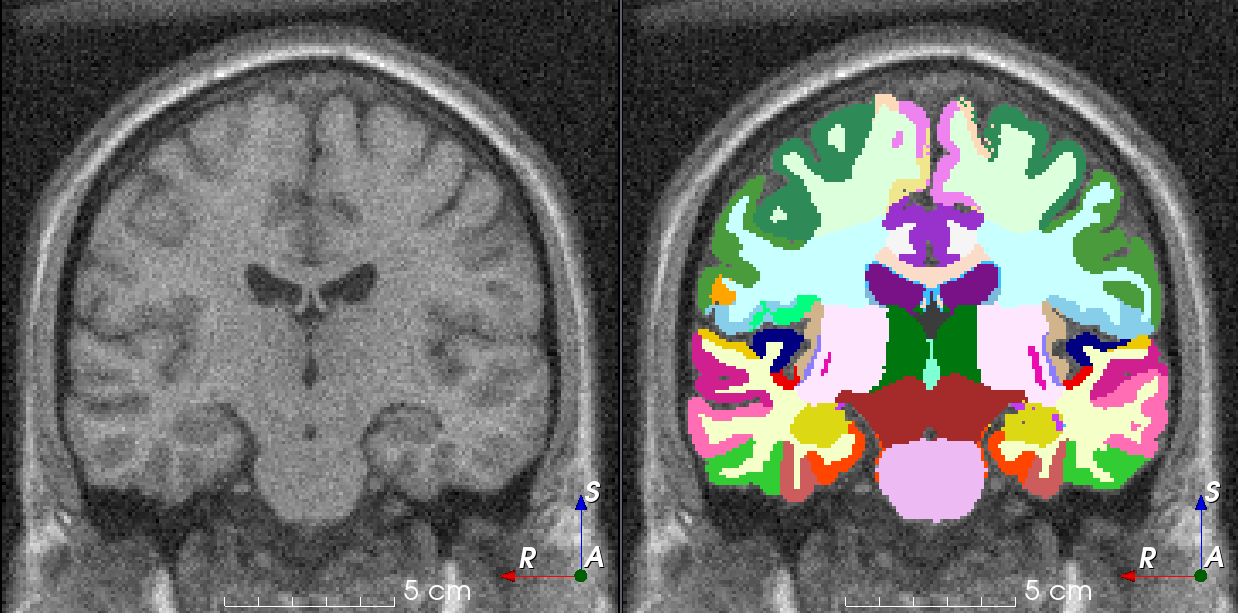}
        \caption{Random noise}
        \label{fig:rnoise}
    \end{subfigure}

    \begin{subfigure}{0.45\textwidth}
        \includegraphics[width=0.98\linewidth]{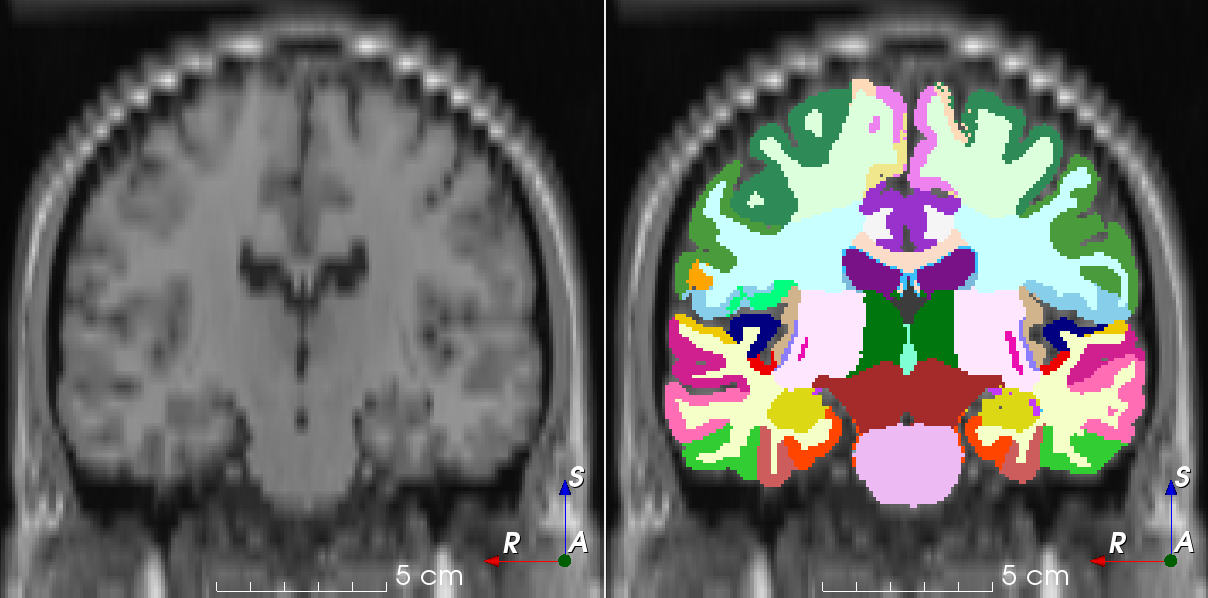}
        \caption{Random anisotropy}
        \label{fig:raffine}
    \end{subfigure}%
    \hspace{2em}%
    \begin{subfigure}{0.45\textwidth}
        \includegraphics[width=0.98\linewidth]{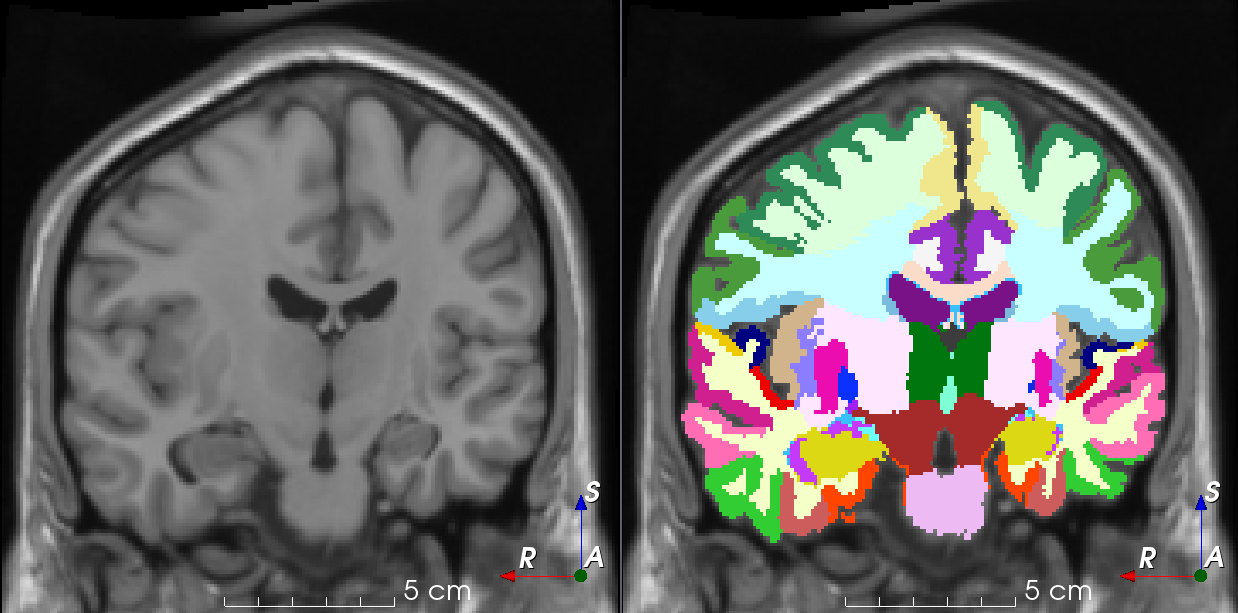}
        \caption{Random elastic transformation}
        \label{fig:relastic}
    \end{subfigure}

    \begin{subfigure}{0.45\textwidth}
        \includegraphics[width=0.98\linewidth]{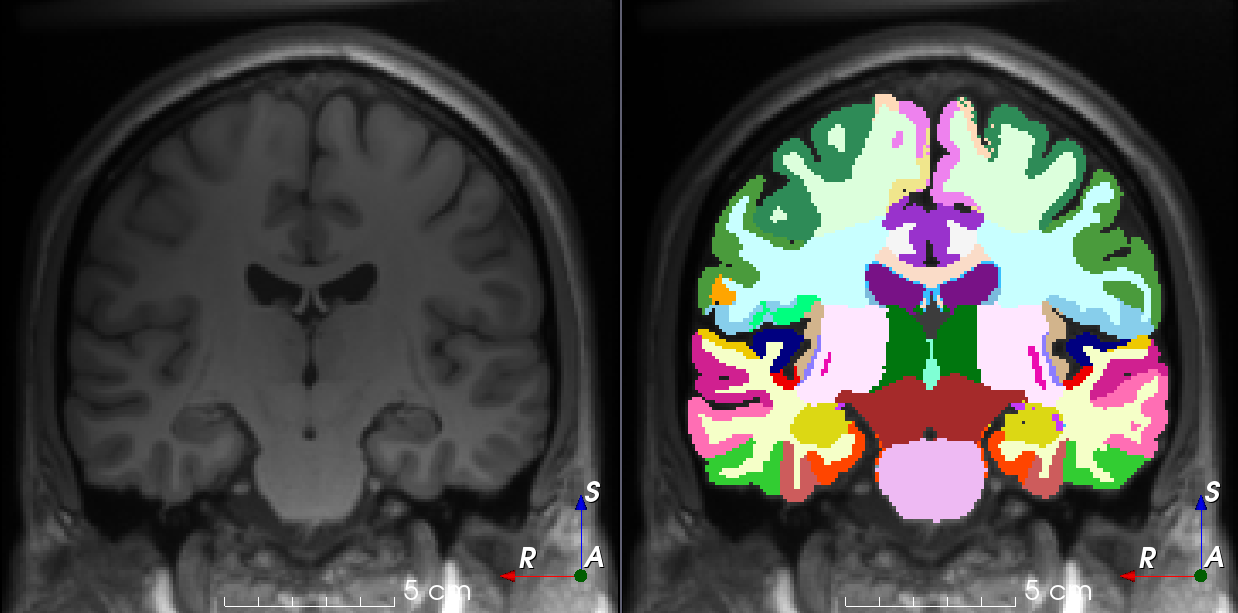}
        \caption{Random bias field artifact}
        \label{fig:rbias}
    \end{subfigure}%
    \hspace{2em}%
    \begin{subfigure}{0.45\textwidth}
        \includegraphics[width=0.98\linewidth]{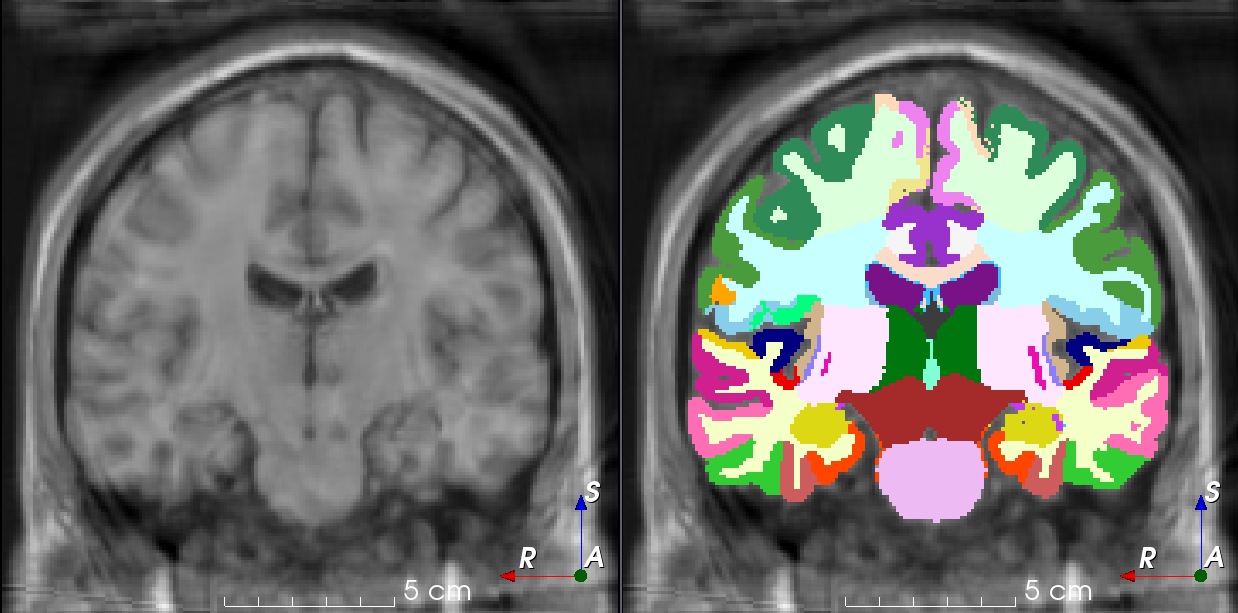}
        \caption{Random motion artifact}
        \label{fig:rmotion}
    \end{subfigure}

    \begin{subfigure}{0.45\textwidth}
        \includegraphics[width=0.98\linewidth]{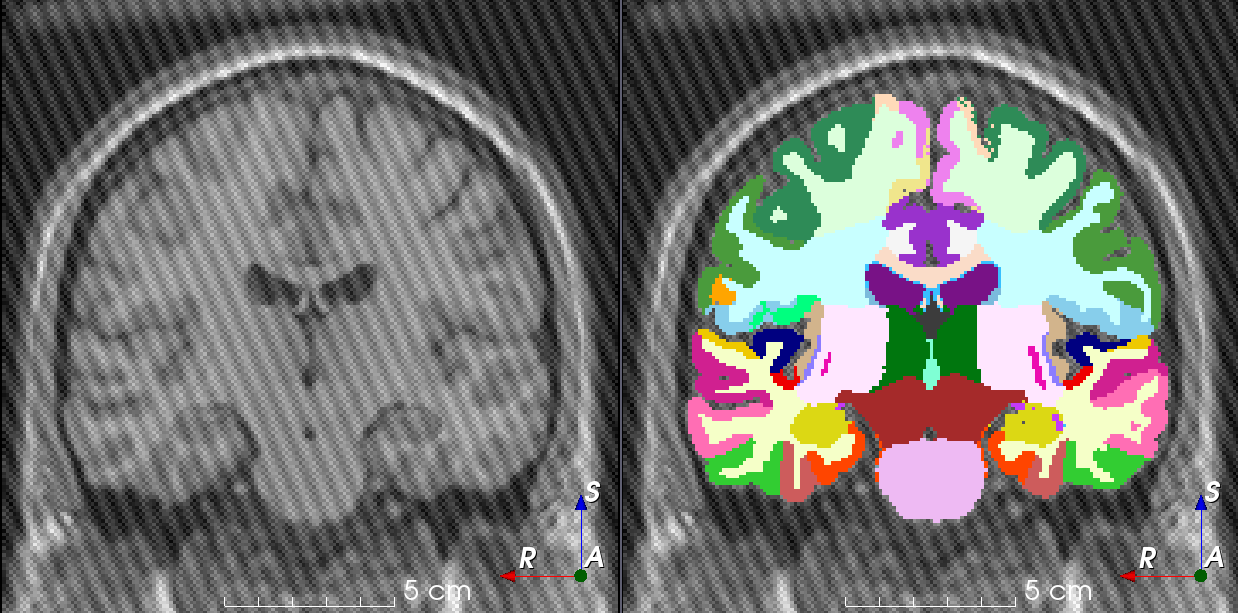}
        \caption{Random spike artifact}
        \label{fig:rspike}
    \end{subfigure}%
    \hspace{2em}%
    \begin{subfigure}{0.45\textwidth}
        \includegraphics[width=0.98\linewidth]{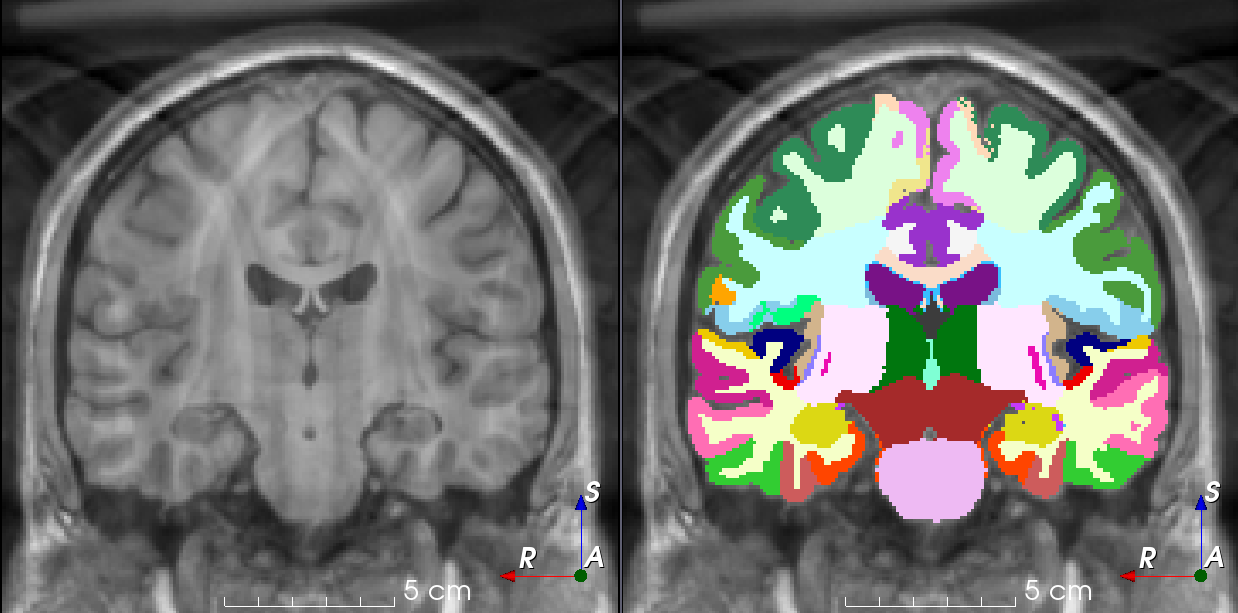}
        \caption{Random ghosting artifact}
        \label{fig:rghost}
    \end{subfigure}

    \caption{%
        A selection of data augmentation techniques
        available in TorchIO \torchioversion.
        Each example is presented as a pair of images composed of the
        transformed image and a corresponding transformed label map.
        Note that all screenshots are from a 2D coronal slice
        of the transformed 3D images.
        The \gls{mri} corresponds to the \gls{mni} Colin 27
        average brain~\citep{holmes_enhancement_1998}, which can
        be downloaded using \texttt{torchio.datasets.Colin27}.
        Label maps were generated using an automated brain parcellation
        algorithm~\citep{cardoso_geodesic_2015}.
    }
    \label{fig:augmentations}
\end{figure}

The transforms \gls{api} was designed to be similar to the PyTorch
\linebreak
\texttt{torchvision.transforms} module.
TorchIO includes augmentations such as
random affine transformation (\cref{fig:raffine})
or random blur (\cref{fig:rblur}),
but they are implemented using medical imaging
libraries~\citep{lowekamp_design_2013,brett_nipynibabel_2020}
to take into account specific properties of medical images,
namely their size, resolution, location, and orientation (see \cref{sec:metadata}).
\cref{tab:transforms} shows transforms implemented in TorchIO \torchioversion
and their main corresponding library dependencies.

\begin{table}[ht]
    \caption{
        Transforms included in TorchIO \torchioversion.
        Logos indicate the main library used to process the images.
        \protect\logo{nipy}:~NiBabel\citep{brett_nipynibabel_2020};
        \protect\logo{itk}:~SimpleITK\citep{lowekamp_design_2013};
        \protect\logo{numpy}:~NumPy\citep{van_der_walt_numpy_2011};
        \protect\logo{pytorch}:~PyTorch\citep{paszke_pytorch_2019}.
    }
    \footnotesize
    \begin{center}
        \begin{tabular}{c||c|c}
                                                    & \textbf{Spatial}                      & \textbf{Intensity}                                                      \\
            \hline
            \hline
            \multirow{5}{*}{\textbf{Preprocessing}} & \trsfl{ToCanonical}{nipy}             &                                                                         \\
                                                    & \trsfl{Resample}{itk}                 & \trsfl{HistogramStandardization}{numpy}~\citep{nyul_standardizing_1999} \\
                                                    & \trsfl{Crop}{itk}                     & \trsfl{RescaleIntensity}{numpy}                                         \\
                                                    & \trsfl{Pad}{itk}                      & \trsfl{ZNormalization}{pytorch}                                         \\
                                                    & \trsfl{CropOrPad}{itk}                &                                                                         \\
            \hline
            \multirow{7}{*}{\textbf{Augmentation}}  &                                       & \trsfl{RandomMotion}{numpy}~\citep{shaw_mri_2019}                       \\
                                                    &                                       & \trsfl{RandomBiasField}{numpy}~\citep{sudre_longitudinal_2017}          \\
                                                    &                                       & \trsfl{RandomGhosting}{numpy}                                           \\
                                                    & \trsfl{RandomAffine}{itk}             & \trsfl{RandomSpike}{numpy}~\citep{shaw_heteroscedastic_2020}            \\
                                                    & \trsfl{RandomElasticDeformation}{itk} & \trsfl{RandomBlur}{numpy}                                               \\
                                                    & \trsfl{RandomFlip}{pytorch}           & \trsfl{RandomGamma}{pytorch}                                            \\
                                                    &                                       & \trsfl{RandomNoise}{pytorch}                                            \\
                                                    &                                       & \trsfl{RandomSwap}{pytorch}~\citep{chen_self-supervised_2019}           \\
                                                    &                                       & \trsfl{RandomLabelsToImage}{pytorch}~\citep{billot_learning_2020}       \\
                                                    &                                       & \trsfl{RandomAnisotropy}{itk}~\citep{billot_partial_2020}               \\
        \end{tabular}
    \end{center}
    \label{tab:transforms}
\end{table}

Transforms are designed to be flexible regarding input and output types.
Following a duck typing approach,
they can take as input
PyTorch tensors,
SimpleITK images,
NumPy arrays,
Pillow images,
Python dictionaries,
and instances of
\texttt{Subject}
and \texttt{Image},
and will return an output of the same type.

TorchIO transforms can be classified into either spatial and intensity transforms,
or preprocessing and augmentation transforms (\cref{tab:transforms}).
All are subclasses of the \texttt{Transform} base class.
Spatial transforms and intensity transforms are related to the
\texttt{SpatialTransform} and \texttt{IntensityTransform} classes, respectively.
Transforms whose parameters are randomly chosen are subclasses of \texttt{RandomTransform}.

Instances of \texttt{SpatialTransform} typically modify the image bounds or spacing, and often need
to resample the image using interpolation.
They are applied to all image types.
Instances of \texttt{IntensityTransform} do not modify the position of voxels, only their values,
and they are only applied to instances of \texttt{ScalarImage}.
For example, if a \texttt{RandomNoise} transform
(which is a subclass of
\texttt{IntensityTransform})
receives as input
a \texttt{Subject} with a \texttt{ScalarImage} representing
a \gls{ct} scan and a \texttt{LabelMap} representing a segmentation,
it will add noise to only the \gls{ct} scan.
On the other hand, if a \texttt{RandomAffine} transform
(which is a subclass of \texttt{SpatialTransform}) receives the same input,
the same affine transformation will be applied to both images,
with nearest-neighbor interpolation always used to interpolate
\texttt{LabelMap} objects.

\comment{
Preprocessing transforms are typically applied to normalize images during
training and inference, whereas augmentation transforms
are random operations applied during training to artificially increase the
size of the training dataset~\citep{shorten_survey_2019}.
}

\subsubsection{Preprocessing}

Preprocessing transforms are necessary to ensure
spatial and intensity uniformity of training instances.

Spatial preprocessing is important as \glspl{cnn} do not generally take into
account metadata related to medical images (see \cref{sec:metadata}),
therefore it is necessary to ensure that voxels across images have
similar spatial location and relationships before training.
Spatial preprocessing transforms typically used in medical imaging include
resampling (e.g., to make voxel spacing isotropic for all training samples)
and reorientation (e.g., to orient all training samples in
the same way).
For example, the \texttt{Resample} transform can be used to fix the issue
presented in \cref{fig:metadata}.

Intensity normalization is generally beneficial
for optimization of neural networks.
TorchIO provides intensity
normalization techniques including min-max scaling or
standardization\footnote{In this context,
standardization refers to correcting voxel intensity values
to have zero mean and unit variance.},
which are computed using pure PyTorch.
A binary image, such as a mask representing the foreground
or structures of interest,
can be used to define the set of voxels to be taken into account
when computing statistics for intensity normalization.
We also provide a method for \gls{mri} histogram
standardization~\citep{nyul_new_2000}, computed using NumPy,
which may be used to overcome the differences
in intensity distributions between images acquired using
different scanners or sequences.

\glsreset{mni}  

\subsubsection{Augmentation}

TorchIO includes spatial augmentation transforms such as random flipping using
PyTorch and random affine and elastic deformation transforms using SimpleITK.
Intensity augmentation transforms include
random Gaussian blur using a SimpleITK filter (\cref{fig:rblur})
and addition of random Gaussian noise using pure PyTorch (\cref{fig:rnoise}).
All augmentation transforms are subclasses of \texttt{RandomTransform}.

Although current domain-specific data augmentation transforms available
in TorchIO are mostly related to \gls{mri},
we encourage users to contribute physics-based data augmentation
techniques for \gls{us} or \gls{ct}~\citep{omigbodun_effects_2019}.

We provide several \gls{mri}-specific augmentation transforms
related to $k$-space, which are described below.
An MR image is usually reconstructed as the magnitude
of the inverse Fourier transform of the $k$-space signal, which is populated with
the signals generated by the sample
as a response to a radio-frequency electromagnetic pulse.
These signals are modulated using coils that create gradients of the magnetic
field inside the scanner.
Artifacts are created by using $k$-space transforms to perturb the Fourier space
and generate corresponding intensity artifacts in image space.
The forward and inverse Fourier transforms are computed using the
\gls{fft} algorithm implemented in NumPy.

\paragraph{Random $k$-space spike artifact}

Gradients applied at a very high duty cycle may produce bad
data points, or noise spikes, in $k$-space~\citep{zhuo_mr_2006}.
These points in $k$-space generate a spike artifact,
also known as Herringbone, crisscross or corduroy artifact,
which manifests as uniformly-separated stripes in image space,
as shown in \cref{fig:rspike}.
This type of data augmentation has recently been used to estimate uncertainty
through a heteroscedastic noise model~\citep{shaw_heteroscedastic_2020}.

\paragraph{Random $k$-space motion artifact}

The $k$-space is often populated line by line,
and the sample in the scanner is assumed to remain static.
If a patient moves during the \gls{mri} acquisition, motion artifacts will appear
in the reconstructed image.
We implemented a method to simulate random motion artifacts
(\cref{fig:rmotion}) that has been used successfully for data augmentation
to model uncertainty and improve segmentation~\citep{shaw_mri_2019}.

\paragraph{Random $k$-space ghosting artifact}

Organs motion such as respiration or cardiac pulsation may generate ghosting
artifacts along the phase-encoding direction~\citep{zhuo_mr_2006}
(see \cref{fig:rghost}).
We simulate this phenomenon by removing every $n$th plane of the $k$-space along
one direction to generate $n$ ghosts along that dimension,
while keeping the center of $k$-space intact.

\paragraph{Random bias field artifact}

Inhomogeneity of the static magnetic field in the \gls{mri} scanner produces
intensity artifacts of very low spatial frequency
along the entirety of the image.
These artifacts can be simulated using polynomial basis
functions~\citep{van_leemput_automated_1999},
as shown in \cref{fig:rbias}.

\subsubsection{Composability}
\label{sec:composability}

All transforms can be composed in a linear fashion,
as in the PyTorch \texttt{torchvision} library,
or building a \gls{dag}
using the \texttt{OneOf} transform
(as in~\citep{buslaev_albumentations_2020}).
For example, a user might want to apply a random spatial augmentation transform
to $50\%$ of the samples using either an affine or an elastic transform,
but they want the affine transform to be applied to $80\%$ of the augmented
images, as the execution time is faster.
Then, they might want to rescale the volume intensity for all images
to be between 0 and 1.
\Cref{fig:dag} shows a graph representing the transform composition.
This transform composition can be implemented with just three statements:

\begin{figure}
    \centering
    \includegraphics[%
        width=\linewidth,
        trim = {0 1.5cm 0 3cm},
        clip
    ]{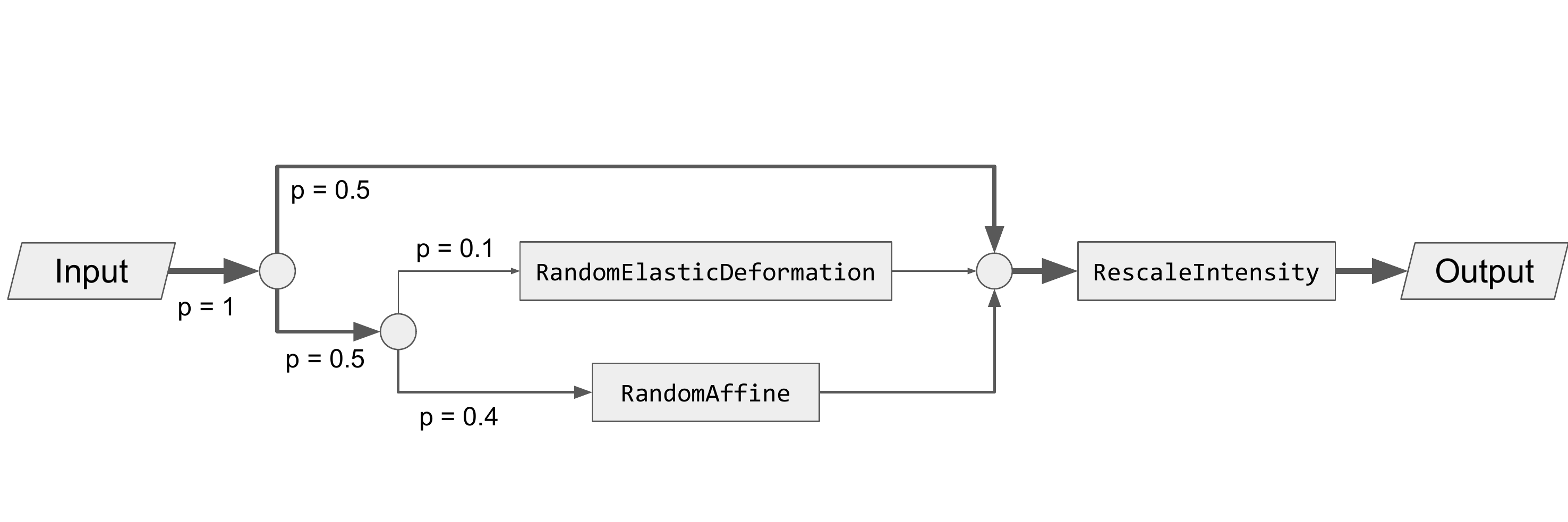}
    \caption{%
        Graph representation of the composed transform
        described in \cref{sec:composability}.
    }
    \label{fig:dag}
\end{figure}

\begin{minted}[
    % mathescape,
    % linenos,
    % numbersep=5pt,
    gobble=2,
    frame=lines,
    framesep=2mm
    ]{python}
  import torchio as tio
  spatial_transforms = {
      tio.RandomElasticDeformation(): 0.2,
      tio.RandomAffine(): 0.8,
  }
  transform = tio.Compose([
      tio.OneOf(spatial_transforms, p=0.5),
      tio.RescaleIntensity((0, 1)),
  ])
\end{minted}

\texttt{Compose} and \texttt{OneOf} are implemented as TorchIO
transforms.

\subsubsection{Extensibility}

The \texttt{Lambda} transform can be passed an arbitrary callable object, which
allows the user to augment the library with custom transforms without having
a deep understanding of the underlying code.

Additionally, more complex transforms can be developed.
For example, we implemented a TorchIO transform to simulate brain resection
cavities from preoperative MR images within a self-supervised learning
pipeline~\cite{perez-garcia_simulation_2020}.
The \texttt{RandomLabelsToImage} transform may be used to simulate an image from
a tissue segmentation.
It can be composed with \texttt{RandomAnisotropy} to train neural networks
agnostic to image contrast and
resolution~\citep{billot_learning_2020, billot_partial_2020,iglesias_joint_2020}.

\subsubsection{Reproducibility and traceability}

To promote open science principles, we designed TorchIO to support experiment
reproducibility and traceability.

All transforms support receiving Python primitives as arguments,
which makes TorchIO suitable to be used with a configuration file associated to
a specific experiment.

A history of all applied transforms and their computed
random parameters is saved in the transform output so that the path in the
\gls{dag} and the parameters used can be traced and reproduced.
Furthermore, the \texttt{Subject} class includes a method to compose the
transforms history into a single transform that may be used to reproduce the
exact result (\cref{sec:composability}).

\subsubsection{Invertibility}

Inverting transforms is especially useful in scenarios where one needs to
apply some transformation, infer a segmentation on the transformed data and
then apply the inverse transformation to bring the inference into the original image space.
The \texttt{Subject} class includes a method to invert the transformations applied.
It does this by first inverting all transforms that are invertible, discarding the ones that are not.
Then, it composes the invertible transforms into a single transform.

Transforms invertibility is most commonly applied to
test-time augmentation~\citep{moshkov_test-time_2020}
or estimation of aleatoric uncertainty~\citep{wang_aleatoric_2019}
in the context of image segmentation.

    \section{Results}

\subsection{Code availability}

All the code for TorchIO is available on
GitHub\fnurl{https://github.com/fepegar/torchio}.
We follow the semantic versioning system~\cite{preston-werner_semantic_2020}
to tag and release our library.
Releases are published on the Zenodo data
repository\fnurl{https://zenodo.org/} to allow users to cite
the specific version of the package they used in their experiments.
The version described in this paper is
\torchioversion~\citep{perez-garcia_fepegartorchio_2020}.
Detailed \gls{api} documentation is hosted on Read the Docs
and comprehensive Jupyter notebook tutorials are hosted on Google Colaboratory,
where users can run examples online.
The library can be installed with a single line of code on Windows, macOS or
Linux using the \gls{pip} package manager: \texttt{pip install torchio}.

TorchIO has a strong community of users, with more than 900 stars on GitHub and
more than 5000 \gls{pypi} downloads per
month\fnurl{https://pypistats.org/packages/torchio} as of August 2021.

\subsubsection{Additional interfaces}

The provided \gls{cli} tool \texttt{torchio-transform} allows users
to apply a transform to an image file without using Python.
This tool can be used to visualize only the preprocessing and data augmentation
pipelines and aid in experimental design for a given application.
It can also be used in shell scripts to preprocess and augment
datasets in cases where large storage is available and on-the-fly loading
needs to be faster.

\begin{figure}
    \centering
    \includegraphics[width=\linewidth, trim = {0 0 0 0.1cm}, clip]{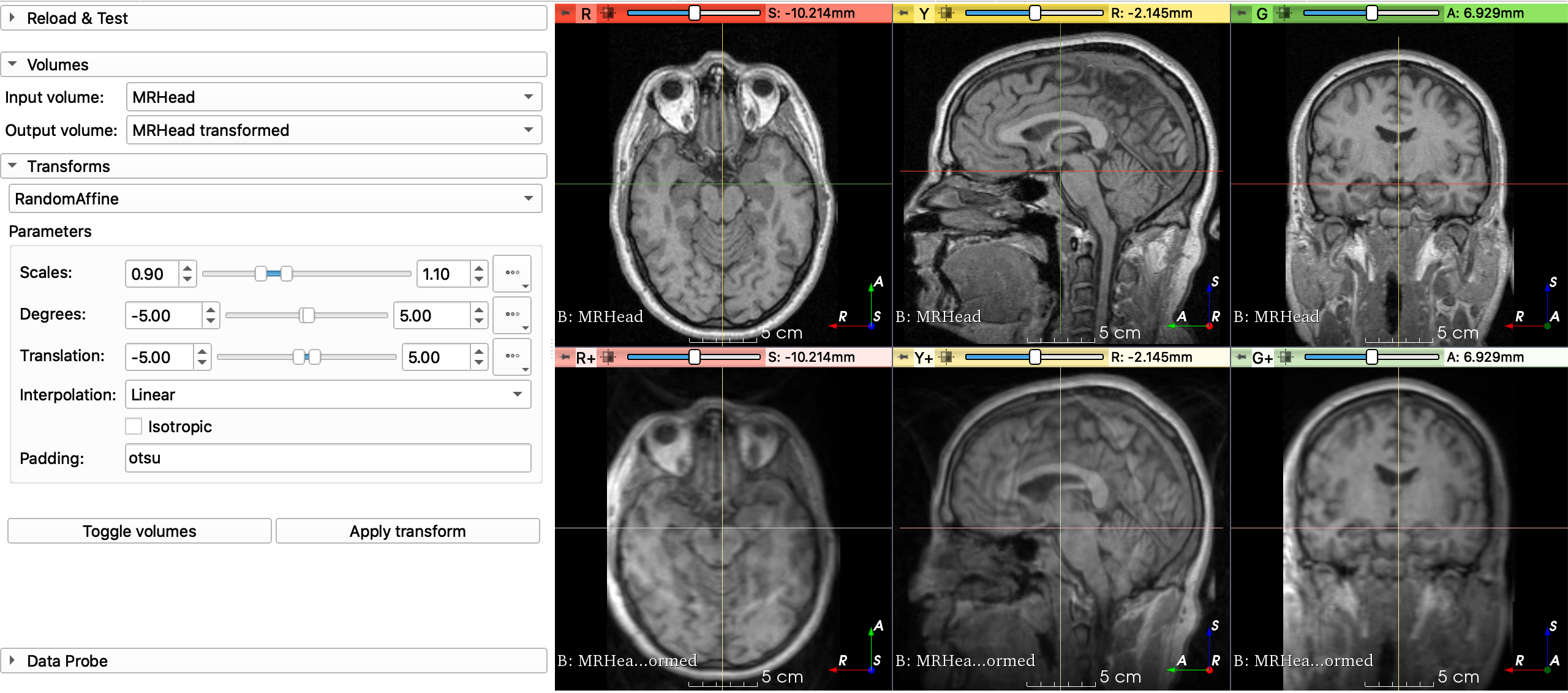}
    \caption{%
        \Gls{gui} for TorchIO,
        implemented as a 3D Slicer extension.
        In this example, the applied transforms are
        \texttt{RandomBiasField},
        \texttt{RandomGhosting},
        \texttt{RandomMotion},
        \texttt{RandomAffine} and
        \texttt{RandomElasticDeformation}.
    }
    \label{fig:slicer}
\end{figure}

Additionally, we provide a \gls{gui} implemented as a Python scripted module
within the \textit{TorchIO} extension available in
3D Slicer~\cite{fedorov_3d_2012}.
It can be used to visualize the effect of the transforms parameters
without any coding (\cref{fig:slicer}).
As with the \gls{cli} tool, users can experimentally assess
preprocessing and data augmentation before network training
to ensure the preprocessing pipeline is suitable for a given application.

    \subsection{Usage examples}

\textcolor{rev2}{%
In this section, we briefly describe the implementations of two medical image
computing papers from the literature, pointing out the TorchIO features that
could be used to replicate their experiments.
}

\subsubsection{Super-resolution and synthesis of MRI}

\textcolor{rev2}{%
In~\citep{iglesias_joint_2020}, a method is proposed to simulate high-resolution
$T_1$-weighted \glspl{mri} from images of different modalities and resolutions.
}

\textcolor{rev2}{%
First, brain regions are segmented on publicly available datasets of brain
\gls{mri}.
During training, an \gls{mri} (\texttt{ScalarImage})
and the corresponding segmentation (\texttt{LabelMap})
corresponding to a specific subject (\texttt{Subject})
are sampled from the training dataset (\texttt{SubjectsDataset}).
Next, the same spatial augmentation transform is applied to both images
by composing an affine transform (\texttt{RandomAffine}) and a nonlinear
diffeomorphic transform (\texttt{RandomElasticDeformation}).
Then, a \gls{gmm} conditioned on the labels is sampled at each voxel location
to simulate an \gls{mri} of arbitrary
contrast (\texttt{RandomLabelsToImage})~\citep{billot_learning_2020}.
Finally, multiple degrading phenomena are simulated on the synthetic image:
variability in the coordinate frames (\texttt{RandomAffine}),
bias field inhomogeneities (\texttt{RandomBiasField}),
partial-volume effects due to a large slice thickness
during acquisition~\citep{billot_partial_2020} (\texttt{RandomAnisotropy}),
registration errors (\texttt{RandomAffine}),
and resampling artifacts (\texttt{Resample}).
}

\subsubsection{Adaptive sampling for segmentation of CT scans}

\textcolor{rev2}{%
In~\citep{berger_adaptive_2018}, \gls{ct} scans that are too large to fit
on a \gls{gpu} are segmented using
patch-based training with weighted sampling of patches.
Discrepancies between labels and predictions are used to create
error maps and patches are preferentially sampled
from voxels with larger error.
}

\textcolor{rev2}{%
During training, a CT scan (\texttt{ScalarImage})
and its corresponding segmentation (\texttt{LabelMap})
from a subject (\texttt{Subject})
are loaded and the same augmentation is performed to both
by applying random rotations and scaling (\texttt{RandomAffine}).
Then, voxel intensities are clipped
to $[-1000, 1000]$ (\texttt{RescaleIntensity})
and divided by a constant factor representing
the standard deviation of the dataset (can be implemented with \texttt{Lambda}).
As the \gls{ct} scans are too large to fit in the \gls{gpu},
patch-based training is used (\texttt{Queue}).
To obtain high-resolution predictions and a large receptive field simultaneously,
two patches of similar size but different \gls{fov} are generated
from each sampled patch:
a context patch generated by downsampling the original patch (\texttt{Resample})
and a full-resolution patch with a smaller \gls{fov} (\texttt{CropOrPad}).
At the end of each epoch, error maps for each subject (\texttt{Subject})
are computed as the difference between the labels and predictions.
The error maps are used in the following epoch to sample patches with large
errors more often (\texttt{WeightedSampler}).
At inference time, a sliding window (\texttt{GridSampler}) is used to predict
the segmentation patch by patch, and patches are aggregated to build the
prediction for the whole input volume (\texttt{GridAggregator}).
}

\newcommand{\remove}[1]{\textcolor{red}{\st{#1}}}

    \section{Discussion}

We have presented TorchIO, a new library to efficiently load, preprocess,
augment and sample medical imaging data during the training of \glspl{cnn}.
It is designed in the style of the deep learning framework PyTorch
to provide medical imaging specific preprocessing and data augmentation
algorithms.

The main motivation for developing TorchIO as an open-source toolkit is to help
researchers standardize medical image processing pipelines and allow them to
focus on the deep learning experiments.
It also encourages good open-science practices, as it supports experiment
reproducibility and is version-controlled so that the software can be cited
precisely.

The library is compatible with other higher-level deep learning frameworks for
medical imaging such as \gls{monai}.
For example, users can benefit from TorchIO's \gls{mri} transforms and
patch-based sampling while using \gls{monai}'s networks, losses, training pipelines
and evaluation metrics.

\textcolor{rev2}{%
The main limitation of TorchIO is that most transforms are not differentiable.
The reason is that PyTorch tensors stored in TorchIO data
structures must be converted to SimpleITK images or NumPy arrays
within most transforms, making them not compatible
with PyTorch's automatic differentiation engine.
However, compatibility between PyTorch and ITK has recently been
improved, partly thanks to the appearance of the \gls{monai}
project~\citep{mccormick_itk_2021}.
Therefore, TorchIO might provide differentiable transforms in the future, which
could be used to implement, e.g., spatial transformer networks for image
registration~\citep{lee_image-and-spatial_2019}.
Another limitation is that many more transforms that are \gls{mri}-specific exist than for other imaging modalities such as \gls{ct} or \gls{us}.
This is in part due to more users working on \gls{mri} applications and requesting \gls{mri}-specific transforms.
However, we welcome contributions for other modalities as well. 
}

In the future, we will work on extending the preprocessing and augmentation
transforms to different medical imaging modalities
such as \gls{ct} or \gls{us}, and improving compatibility with related works.
The source code, as well as examples and documentation,
are made publicly available online, on GitHub.
We welcome feedback, feature requests, and contributions to the library,
either by creating issues on the GitHub repository or by emailing the authors.



\section*{Acknowledgments}

We would like to acknowledge all of the contributors to the TorchIO
library.
We thank the NiftyNet team for their support, and Alejandro Granados, Romain
Valabregue, Fabien Girka, Ghiles Reguig, David Völgyes and Reuben Dorent for
their valuable insight and contributions.

This work is supported by the Engineering and Physical Sciences Research Council (EPSRC) [EP/R512400/1].
This work is additionally supported by the EPSRC-funded UCL Centre for Doctoral Training in Intelligent, Integrated Imaging in Healthcare (i4health) [EP/S021930/1] and the Wellcome / EPSRC Centre for Interventional and Surgical Sciences (WEISS, UCL) [203145Z/16/Z]. %
This publication represents, in part, independent research commissioned by the Wellcome Innovator Award [218380/Z/19/Z/].
The views expressed in this publication are those of the authors and not necessarily those of the Wellcome Trust.

    \bibliography{TorchIO}

\end{document}